\documentclass[12pt]{article}

\usepackage{amsmath}
\usepackage{amssymb}
\usepackage{euscript}

\usepackage{epsfig}

\usepackage{cite}

\usepackage{alltt}

\def\Order#1{${\cal O}(#1$)}
\def\oal{${\cal O}(\alpha)$}
\def\gmu{$G_{\mu}$}

\newcommand{\beq}{\begin{equation}}
\newcommand{\eeq}{\end{equation}}

\begin{document}                     

\allowdisplaybreaks

\begin{titlepage}

\begin{flushright}
{\bf  CERN-TH/2003-018
}
\end{flushright}

\vspace{5mm}
\begin{center}
{\LARGE
Multiphoton Radiation in Leptonic $W$-Boson Decays$^{\star}$
}
\end{center}

\vspace{1mm}
\begin{center}
{\large\bf W. P\l{}aczek$^{a,b}$ 
        {~\rm and~}  
     S. Jadach$^{c,b}$  
}

\vspace{4mm}
{\em
$^a$Institute of Computer Science, Jagellonian University,\\
   ul. Nawojki 11, 30-072 Cracow, Poland,\\ \vspace{1mm}
$^b$CERN, TH Division, CH-1211 Geneva 23, Switzerland,\\ \vspace{1mm}
$^c$Institute of Nuclear Physics,\\
  ul. Radzikowskiego 152, 31-342 Cracow, Poland,\\
}
\end{center}

\vspace{15mm}
\begin{abstract}
We present the calculation of multiphoton radiation effects in 
leptonic $W$-boson decays in the framework of the Yennie--Frautschi--Suura
exclusive exponentiation. This calculation is implemented in the
Monte Carlo event generator WINHAC for single $W$-boson production in
hadronic collisions at the parton level. 
Some numerical results obtained with the help of this program are also 
presented. 
\end{abstract}

\vspace{25mm}
\begin{center}
{\it To be submitted to European Physical Journal C}
\end{center}

\vspace{25mm}
\footnoterule
\noindent
{\footnotesize
\begin{itemize}
\item[${}^{\star}$]
  Work partly supported by 
  the Polish Government grants
  KBN 2P03B00122 
  and KBN 5P03B09320, 
  the EC FP5 contract HPRN-CT-2000-00149, 
  the EC FP5 Centre of Excellence ``COPIRA'' under the contract 
  No.\ IST-2001-37259. 
\end{itemize}
}

\vspace{1mm}
\begin{flushleft}
{\bf CERN-TH/2003-018
\\  January 2003
}
\end{flushleft}

\end{titlepage}

\section{Introduction}

Studying the $W$-boson physics is an important way of testing the Standard 
Model (SM) and searching for ``new physics''. It can be done in both the
electron--positron and hadron colliders. In $e^+e^-$ collisions, the main
source of $W$ bosons is the process of $W$-pair production. This process
was one of the most important subjects of the LEP2 experiments at CERN,
run in the years 1996--2000, see e.g.\ Refs.~\cite{LEP2YR:1996,LEP2YR:2000}.
It also belongs to the main topics of a research programme of future
linear colliders (LC), see e.g.\ Ref.~\cite{TESLA-TDR:2001}. In this 
process one can measure precisely the $W$-boson mass and width as well as
non-abelian triple and quartic gauge-boson couplings%
  \footnote{Actually, for the quartic gauge-boson coupling an additional
            gauge boson, $\gamma$ or $Z$, is required.}. 
In hadron colliders (proton--proton or proton--antiproton), 
the main source of $W$ bosons is the process of single-$W$ production.
The most precise measurements of the $W$-boson mass and width in hadron
colliders come from this process, see e.g.~Ref.~\cite{LHCYR:2000}.
It can also be used to extract parton distribution functions (PDFs)
and to measure parton luminosities~\cite{Dittmar:1997}.

Among radiative corrections that affect the $W$-boson observables 
considerably is the photon radiation in leptonic $W$ decays. It distorts 
$W$-invariant-mass distributions reconstructed from $W$-decay products
in  $e^+e^-$ experiments~\cite{Beenakker:1998cu,yfsww3:1998b} 
or $W$-transverse-mass distributions obtained in hadron-collider 
experiments~\cite{Baur:1998}. 
These distortions are strongly 
acceptance-dependent, see e.g.~Refs.~\cite{Baur:1998,yfsww3:1998b}. 
This radiation also affects lepton pseudorapidity distribution,
which is the main tool for the PDFs and parton luminosities measurements
in the hadron colliders. Therefore, precise theoretical predictions for
the photon radiation in the leptonic $W$ decays is of great importance
for both types of high-energy particle colliders.
In order to be fully applicable in a realistic experimental situation,
such predictions have to be provided in terms of a Monte Carlo
event generator (MCEG). 

The \oal\ electroweak (EW) radiative corrections in the on-shell $W$ decays
were calculated analytically a long time ago by several 
authors~\cite{Marciano:1973,Fleischer:1985,Bardin:1986,Denner:1990}.
In the case of the $W$-pair production in the $e^+e^-$ colliders, the 
calculations of the \oal\ EW corrections in a double-pole approximation (DPA)
were done in Refs.~\cite{Beenakker:1998gr} and \cite{Denner:2000bj}.
The latter were implemented in the MCEG RacoonWW~\cite{Denner:2000bj}.
For single-$W$ production in hadronic collisions, the respective
\oal\ EW corrections were calculated in 
Refs.~\cite{Wackeroth:1996,Baur:1998,Zykunov:2000,Dittmaier:2001ay}.
The MCEG for this process, including pure QED \oal\ corrections, was provided
long ago by Berends and Kleiss~\cite{Berends:1985}.
A two-real-photon radiation cross section in $W$ decays was calculated in
Ref.~\cite{Baur:1999}. 
On the other hand, the MC package PHOTOS~\cite{photos:1994} provides
a universal tool for the generation of photon radiation in particle decays
up to \Order{\alpha^2} in the leading-log (LL) approximation. It was
used in the MCEG YFSWW~\cite{yfsww3:2001} for the simulation of radiative
$W$ decays for the $W$-pair production process in $e^+e^-$ collisions.

To date, however, none of the existing MCEG for $W$-boson physics 
included multiphoton radiation in leptonic $W$ decays through 
exclusive QED exponentiation. Therefore, the influence of higher-order
radiative corrections on the $W$-boson observables was difficult to
assess\footnote{%
                 The calculation of Ref.~\cite{Baur:1999} requires
                 two visible photons in a detector.}.   
In this paper, we provide the first calculation of the multiphoton
radiation in leptonic $W$ decays in the framework of the
Yennie--Frautschi--Suura exclusive exponentiation~\cite{yfs:1961}.
This calculation is implemented in the MCEG for single-$W$ production in 
quark--antiquark collisions called WINHAC~\cite{WINHAC:2002}. 
It is a starting point for the full MC
program for Drell--Yan-like single-$W$ production at the proton--antiproton 
(Tevatron) and proton--proton (LHC) colliders. 
The presented calculation as well
as the respective MC algorithm can also be implemented in the MCEG for
$W$-pair production in the $e^+e^-$ collisions, such as YFSWW.

The paper is organized as follows.  In Section~2 we provide spin
amplitudes for the Born-level process and for the process with single-photon
radiation in $W$ decays. In Section~3 we discuss the YFS exponentiation
in leptonic $W$-boson decays. Numerical results are presented in Section~4.
Section~5 summarizes the paper and gives some outlook.
Finally, the appendices contain supplementary formulae.  

\section{Spin amplitudes}
\label{sec:sa}

In the calculation of matrix elements for the process of single-$W$ production
in hadronic collisions, we use the spin amplitude formalism of 
Ref.~\cite{hagiwara:1986}. In this approach, spinors are expressed
in the Weyl basis, the vector-boson polarizations in the Cartesian basis,
and the spin amplitudes are evaluated numerically for arbitrary four-momenta 
and masses of fermions and bosons. This evaluation amounts, in practice, 
to multiplying 2$\times$2 $c$-number matrices by 2-dimensional $c$-number
vectors. We give below the general spin amplitudes for arbitrary fermions 
in the initial and in the final states, and apply them later on
to the single-$W$ production in $q\bar{q}$ collisions with leptonic 
$W$ decays.

\subsection{Born level}
\label{ssec:born}

\begin{figure}[!ht]
\centering
\epsfig{file=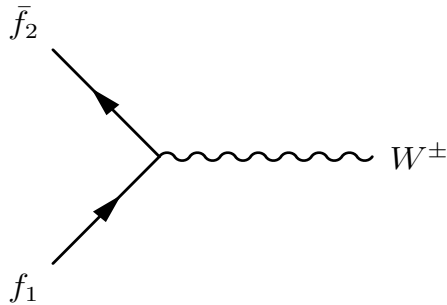,width=60mm,height=40mm}
\caption{\sf
 The Born-level Feynman diagram for single-$W$ production 
 in fermion--antifermion collisions.
}
\label{fig:Wprod}
\end{figure}
The Born-level Feynman diagram for single-$W$ production 
in fermion--antifermion collisions
\begin{equation}
f_1(p_1,\sigma_1) + \bar{f_2}(p_2,\sigma_2) \longrightarrow W^{\pm}(Q,\lambda)
\label{eq:Wprod}
\end{equation}
is depicted in Fig.~\ref{fig:Wprod}, where $(p_i,\sigma_i)$ denotes the
four-momentum and helicity ($\sigma_i=\pm 1$) of the corresponding fermion, 
while $(Q,\lambda)$ is the four-momentum and polarization of the $W$-boson
($\lambda=1,2,3$). The fermions $f_1$ and $f_2$ are members of 
$SU(2)_L$ doublets with opposite values of the weak-isospin third component
and the pair $f_1\bar{f_2}$ is the $SU(3)_c$ singlet.
The spin amplitudes for this process, in the convention
of Ref.~\cite{hagiwara:1986}, read
\begin{equation}
{\cal M}_P^{(0)}(\sigma_1,\sigma_2;\lambda) = 
-\frac{ieV_{f_1 f_2}}{\sqrt{2}s_W}\,
 \omega_{-\sigma_1}(p_1)\,\omega_{\sigma_2}(p_2)\,\sigma_2\,
 S\left(p_2,\epsilon_W^{\ast}(Q,\lambda),p_1\right)_{-\sigma_2,\sigma_1}^-,
\label{eq:saWp}
\end{equation}
where $e$ is the positron electric charge, $V_{f_1 f_2}$ is the element
of the weak-mixing matrix (the CKM matrix for quarks, 
the MNS matrix for leptons%
    \footnote{In the following we neglect masses of neutrinos and therefore
              do not consider mixing in the lepton sector.}%
), $s_W=\sin\theta_W$, with $\theta_W$ the weak-mixing (Weinberg) angle;
\begin{equation}
\omega_{\pm}(p) = \sqrt{p^0 \pm |\vec{p}|}\,;
\label{eq:omega}
\end{equation}
$\epsilon_W(Q,\lambda)$ is the $W$-boson polarization vector ($\ast$ denotes
the $c$-number conjugation);
and $S(\ldots)$ is the spinorial string function, given explicitly
in Appendix~A. The above spin amplitudes are identical for any colour singlet
of the initial fermion pair $f_1\bar{f_2}$.

\begin{figure}[!ht]
\centering
\epsfig{file=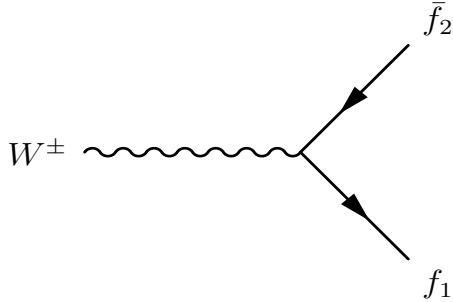,width=60mm,height=40mm}
\caption{\sf
 The Born-level Feynman diagram for $W$-boson decay. 
}
\label{fig:Wdec0}
\end{figure}
The spin amplitudes for the Born-level $W$-boson decay:
\begin{equation}
W^{\pm}(Q,\lambda) \longrightarrow f_1(q_1,\tau_1) + \bar{f_2}(q_2,\tau_2),
\label{eq:Wdec0}
\end{equation}
shown diagrammatically in Fig.~\ref{fig:Wdec0}, are given by
\begin{equation}
{\cal M}_D^{(0)}(\lambda;\tau_1,\tau_2) = -\frac{ieCV_{f_1 f_2}}{\sqrt{2}s_W}\,
 \omega_{-\tau_1}(q_1)\,\omega_{\tau_2}(q_2)\,\tau_2\,
 S\left(q_1,\epsilon_W(Q,\lambda),q_2\right)_{\tau_1,-\tau_2}^-,
\label{eq:saWd0}
\end{equation}
where $\tau_{1,2}$ denote the helicities of the final-state fermions,
and $C$ is the colour factor
\begin{equation}
 C =
\left\{
 \begin{array}{l}
    \sqrt{3}: \hspace{2mm} \mathrm{for\ quarks,}\\
    1: \hspace{5.5mm}
     \mathrm{\rm for\ leptons.}
  \end{array}
\right.
\label{eq:Colfac}
\end{equation}

The above spin amplitudes can be easily translated from the vector-boson
Cartesian basis into the helicity basis, using the following 
transformations:
\begin{equation}
\begin{aligned}
  {\cal M}_{hel}(\lambda=\pm) & = \frac{1}{\sqrt{2}}\,
    \left[\,\mp{\cal M}(\lambda=1) 
    - i{\cal M}(\lambda=2)\,\right]\,, \\
   {\cal M}_{hel}(\lambda=0) & = {\cal M}(\lambda=3).
\end{aligned}
\label{eq:sahel}
\end{equation}

Then, the Born-level matrix element for the single-$W$ production and decay
is given by the coherent sum of the above
spin amplitudes over the $W$-boson polarizations multiplied by the 
Breit-Wigner function corresponding to the $W$ propagator:
\begin{equation}
{\cal M}^{(0)}(\sigma_1,\sigma_2;\tau_1,\tau_2)
 \, = \, \frac{1}{Q^2 - M_W^2 + i\gamma_W(Q^2)} \,
 \sum_{\lambda}
 {\cal M}_P^{(0)}(\sigma_1,\sigma_2;\lambda)\, 
 {\cal M}_D^{(0)}(\lambda;\tau_1,\tau_2),  
\label{eq:saW0}
\end{equation}
where 
\begin{equation}
 \gamma_W(Q^2) =
\left\{
 \begin{array}{l}
    M_W\Gamma_W: \hspace{9mm}     \text{in the fixed-width scheme,}\\
    Q^2\Gamma_W/M_W: \hspace{2mm} \text{in the running-width scheme.}
  \end{array}
\right.
\label{eq:GamW}
\end{equation}
It is known that the fixed-width and running-width schemes are connected by
an appropriate rescaling of the line-shape parameters, here $M_W$ and
$\Gamma_W$~\cite{Bardin:1988}. 

\subsection{\oal\ corrections}
\label{ssec:1ord}

The cross section for Drell--Yan-like $W$ production in hadronic collisions is
dominated by the resonant single-$W$ process. Therefore, it can be 
described to a good accuracy with the help of the leading-pole approximation
(LPA)~\cite{Wackeroth:1996, Dittmaier:2001ay}. The non-LPA contributions 
are important only for specific high-$W$-invariant-mass observables 
(e.g.\ in ``new physics'' searches). In this paper we concentrate on the
resonant $W$ production; the non-resonant contributions will be included
later on.  
The \oal\ EW radiative corrections to the resonant single-$W$ production 
and decay can be divided in a gauge-invariant way into the initial-state 
corrections (ISR), initial--final interferences (non-factorizable corrections) 
and the final-state corrections (FSR), see 
e.g.\ Refs.~\cite{Wackeroth:1996,Dittmaier:2001ay}.
The leading ISR (mass-singular) QED corrections  can be absorbed in the parton 
distribution functions, in a way similar to the leading QCD 
corrections~\cite{Baur:1998,Haywood:2000,Dittmaier:2001ay}. 
In general, the ISR corrections 
have a rather minor effect on the single-$W$ observables at hadron 
colliders~\cite{Baur:1998,Haywood:2000}. 
The non-factorizable corrections are negligible in resonant $W$-boson 
production~\cite{Wackeroth:1996,Baur:1998}. On the contrary, the FSR 
corrections affect various $W$ observables considerably~\cite{Baur:1998}.
This paper is devoted to the FSR, and the other corrections will
be considered in the future. 
More precisely, our aim here is to give a theoretical description
of the QED part of the FSR corrections in the framework of the YFS 
exclusive exponentiation. 

It is known that in processes involving the $W$-bosons, 
the electroweak corrections cannot be split in 
a gauge-invariant way into the pure-QED and pure-weak ones. 
However, one can extract some parts of photonic corrections that are 
gauge-independent, see e.g.\ Refs.~\cite{Marciano:1973,Wackeroth:1996}.
In this paper we follow the approach of Ref.~\cite{Marciano:1973},
where only the infrared-singular and fermion-mass-logarithmic terms are 
extracted from the virtual \oal\ EW corrections and combined with 
the real-photon contributions. They form the so-called QED-like corrections.
The rest of the virtual photonic corrections can be combined with 
the genuine weak-boson corrections to form the so-called weak-like corrections.
Another solution, based on the YFS separation of the infrared (IR) QED terms,
was presented in Ref.~\cite{Wackeroth:1996}. It differs from the previous
one by subleading (non-log) terms. It can also be easily implemented
in our calculations. In this approach, however, the weak-like corrections 
are slightly larger numerically. Of course, when the whole \oal\ EW
corrections are included these two approaches are equivalent. 
Since in this paper we deal with QED-like corrections only, we have chosen
the solution of  Ref.~\cite{Marciano:1973}, which is closer to the full 
\oal\ calculation. This, however, may change in the future when also 
the weak-like corrections are included.

The major portion of the electroweak corrections can be taken into account 
by using the so-called \gmu\ scheme, i.e. parametrizing the 
cross section by the Fermi constant \gmu\ 
instead of the fine-structure constant $\alpha$, 
see e.g.\ Refs.~\cite{fleischer:1989,Dittmaier:2001ay}. 
In our case, this amounts to the replacement
\begin{equation}
\alpha = \frac{e^2}{4\pi}\: \longrightarrow \: \alpha_{G_{\mu}} = 
      \frac{\sqrt{2} G_{\mu} M_W^2 s_W^2}{\pi}
\label{eq:Gmu-scheme}
\end{equation}
in the hard-process parts of the matrix elements.

\subsubsection{Virtual and real soft-photon corrections}
\label{ssec:virsof}

The virtual QED-like correction to the leptonic $W$-boson decay, extracted
from  Ref.~\cite{Marciano:1973}, reads
\begin{equation}
\delta^v_{\rm QED}(M,m_l) = \frac{\alpha}{\pi}
  \left[2 \left( \ln\frac{M}{m_l} - 1 \right)\ln\frac{m_{\gamma}}{M} 
   + \ln^2\frac{M}{m_l} + \frac{1}{2}\ln\frac{M}{m_l} \right], 
\label{eq:vir1}
\end{equation}
where $M$ is the $W$ invariant mass (i.e. $M^2=Q^2$), 
$m_l$ is the charged-lepton mass
and $m_{\gamma}$ a dummy photon mass (an IR regulator). After combining 
the virtual correction
with the real-soft-photon contribution, one obtains the 
virtual$\,+\,$real-soft-photon correction
(cf.\ e.g.\ Refs.~\cite{Marciano:1973,Fleischer:1985}):
\begin{equation}
\delta^{v+s}_{\rm QED}(M,m_l) = \frac{\alpha}{\pi}
  \left[2 \left(\ln\frac{M}{m_l} - 1\right)\ln\frac{2k_s}{M} 
   + \frac{3}{2}\ln\frac{M}{m_l} - \frac{\pi^2}{6} + 1 \right], 
\label{eq:virsof1}
\end{equation}
where $k_s$ is the soft-photon cut-off, i.e.\ the maximum energy of the
soft real photon up to which its contribution has been integrated over.
When the above correction is combined with the appropriate  real-hard-photon 
contribution integrated over the remaining photon phase space, 
one obtains the total QED-like correction to the $W$-boson
width~\cite{Marciano:1973,Berends:1985}
\begin{equation}
\delta^{tot}_{\rm QED} = 
\frac{\alpha}{\pi}\left(\frac{77}{24} - \frac{\pi^2}{3} \right)
\simeq -1.89\times 10^{-4},
\label{eq:QED1}
\end{equation}
which does not contain mass-logarithmic terms, 
in accordance with the KLN-theorem~\cite{Kinoshita:1962,LeeNau:1964},
and is small numerically.
The above formulae were obtained in the small-lepton-mass approximation,
$m_l \ll M$, which means that the terms ${\cal O}(m_l^2/M^2)$ were neglected.

\subsubsection{Real hard-photon radiation}
\label{ssec:hard1}

\begin{figure}[!ht]
\centering
\epsfig{file=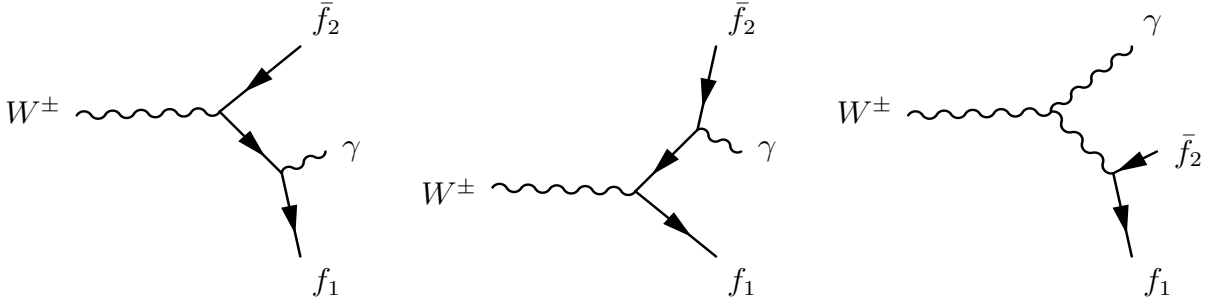,width=160mm,height=40mm}
\caption{\sf
 The Feynman diagrams 
 for $W$-boson decay including single real-photon radiation 
 (in the unitary gauge).  
}
\label{fig:Wdec1}
\end{figure}
Here we present the scattering amplitudes for single hard-photon radiation
in leptonic $W$-boson decays using the spin-amplitude formalism of
Ref.~\cite{hagiwara:1986} and the notation introduced in the previous
subsections. For the process
\begin{equation}
W^{\pm}(Q,\lambda) \longrightarrow f_1(q_1,\tau_1) + \bar{f_2}(q_2,\tau_2)
                    + \gamma(k,\kappa),
\label{eq:Wdec1}
\end{equation}
given by the Feynman diagrams in Fig.~\ref{fig:Wdec1}, we obtain the \oal\ 
spin amplitudes
\begin{equation}
\begin{aligned}
{\cal M}_D^{(1)} & (\lambda;\tau_1,\tau_2,\kappa) 
  = -\frac{ie^2CV_{f_1 f_2}}{\sqrt{2}s_W}\,
 \omega_{-\tau_1}(q_1)\,\omega_{\tau_2}(q_2)\,\tau_2 \\
 \times \Bigg\{ &
 \left(\frac{Q_{f_1}\, q_1\cdot\epsilon_{\gamma}^{\ast}}{k\cdot q_1}
      -\frac{Q_{f_2}\, q_2\cdot\epsilon_{\gamma}^{\ast}}{k\cdot q_2} 
      -\frac{Q_W\, Q\cdot\epsilon_{\gamma}^{\ast}}{k\cdot Q}\right) 
 S\left(q_1,\epsilon_W(Q,\lambda),q_2\right)_{\tau_1,-\tau_2}^- \\
& +  \frac{Q_{f_1}}{2\,k\cdot q_1}
 S\left(q_1,\epsilon_{\gamma}^{\ast}(k,\kappa),k,
        \epsilon_W(Q,\lambda),q_2\right)_{\tau_1,-\tau_2}^- \\
& - \frac{Q_{f_2}}{2\,k\cdot q_2}
 S\left(q_1,\epsilon_W(Q,\lambda),k,
        \epsilon_{\gamma}^{\ast}(k,\kappa),q_2\right)_{\tau_1,-\tau_2}^-\\
& - \frac{Q_W\,k\cdot\epsilon_W}{2\,k\cdot Q}
 S\left(q_1,\epsilon_{\gamma}^{\ast}(k,\kappa),q_2\right)_{\tau_1,-\tau_2}^-
  + \frac{Q_W\,\epsilon_W\cdot\epsilon_{\gamma}^{\ast}}{2\,k\cdot Q}
 S\left(q_1,k,q_2\right)_{\tau_1,-\tau_2}^-
\Bigg\},
\end{aligned}
\label{eq:saWd1}
\end{equation}
where $\epsilon_{\gamma}(k,\kappa)$ is the $\kappa$th polarization 
vector of the photon with four-momentum $k$ (because the photon
is massless, $\kappa=1,2$);
$Q_{f_1},\,Q_{f_2}$ and $Q_W$ are the electric charges (in units of the
positron charge) of the fermions $f_1,\,f_2$ and the $W$-boson, respectively;
they satisfy the condition: $Q_W = Q_{f_1} - Q_{f_2}$.
The spinorial functions $S(\ldots)$ are given explicitly in Appendix~A.
The QED gauge invariance for these amplitudes means that
\begin{equation}
{\cal M}_D^{(1)}(\epsilon_{\gamma}\rightarrow k) = 0. 
\label{eq:GI}
\end{equation}
We have 
checked numerically that after the replacement 
$\epsilon_{\gamma}\rightarrow k$ in Eq.~(\ref{eq:saWd1}),
the values of the spin amplitudes are consistent with zero
within the double-precision accuracy.

Then, the matrix element for single-$W$ production and radiative
$W$ decay can be obtained through
\begin{equation}
{\cal M}^{(1)}(\sigma_1,\sigma_2;\tau_1,\tau_2,\kappa)
  =  \frac{1}{Q^2 - M_W^2 + i\gamma_W(Q^2)} 
 \sum_{\lambda}
 {\cal M}_P^{(0)}(\sigma_1,\sigma_2;\lambda)\, 
 {\cal M}_D^{(1)}(\lambda;\tau_1,\tau_2,\kappa),  
\label{eq:saW1}
\end{equation}
where the lowest-level spin amplitude ${\cal M}_P^{(0)}$ for the single-$W$
production is given in Eq.~(\ref{eq:saWp}).
This matrix element is a coherent convolution of non-radiative spin 
amplitudes for $W$ production and radiative spin amplitudes for $W$ decay.
This means that it describes the photon radiation in the $W$-decay stage
only. 

As was noticed in Ref.~\cite{Berends:1985}, the matrix element for
the single-photon radiation in the Drell--Yan-like $W$ production
process can be, in the fixed-width scheme, split gauge-invariantly 
into the sum of matrix elements for radiative $W$ production
convoluted with non-radiative $W$ decay and  
non-radiative $W$ production convoluted with radiative $W$ decay. 
This can be achieved by exploiting the partial
fraction decomposition of a product of $W$-boson propagators arising
when photon is emitted from an intermediate $W$-boson line~\cite{Berends:1985}.
This simple decomposition, however, does not work in the running-width
scheme. In this case we have
\begin{equation}
\begin{aligned}
\,&
 \frac{1}{Q^2 - M_W^2 + i\gamma_W(Q^2)}\,\frac{1}{{Q'}^2 - M_W^2 
  + i\gamma_W({Q'}^2)}
 = \\
\, &\bigg[
  \underbrace{\frac{1}{2kQ'}\,\frac{1}{{Q'}^2 - M_W^2 + i\gamma_W({Q'}^2)}
                 }_\mathrm{\leftarrow\, production}
 -  
  \underbrace{\frac{1}{Q^2  - M_W^2 + i\gamma_W(Q^2)}\,\frac{1}{2kQ} 
             }_\mathrm{decay\,\rightarrow}
\bigg] 
\frac{1}{1+i\Gamma_W/M_W},
\end{aligned}
\label{eq:pfdWp}
\end{equation}
where $Q$ and $Q'$ are the $W$-boson four-momenta before and after
the emission of the photon with four-momentum $k$: $Q'=Q-k$.
The two terms in the square brackets correspond to the radiative production
and the radiative decay, respectively, but they are multiplied
by the factor $(1+i\Gamma_W/M_W)^{-1}$. So in the case of the running-width 
scheme, the partial fraction decomposition of the $W$-propagator works
modulo this multiplicative factor. However, including the running
$W$-boson width in the case of the photon radiation off the $W$-line 
leads to a violation of the QED Ward identity, 
see e.g.\ Refs.~\cite {Baur:1995,argyres:1995}.
As was shown in Ref.~\cite{Baur:1995}, in order to restore the respective
Ward identity it is sufficient to include the light-fermion-loop corrections
to the $WW\gamma$ vertex. In the small-fermion-mass approximation this
amounts to multiplying the respective radiative amplitude by the factor
\begin{equation}
G_{\rm FLS} = 1 + i\frac{\Gamma_W}{M_W}.
\label{eq:FLS}
\end{equation}
When we multiply our Eq.~(\ref{eq:pfdWp}) by $G_{\rm FLS}$,
the factor outside the square brackets on the r.h.s. drops out,
and we obtain the decomposition of the corresponding amplitude into the 
radiative production and the radiative decay -- exactly as in the fixed-width 
scheme.
Therefore, our matrix element of Eq.~(\ref{eq:saW1}) for single-$W$
production with radiative decays is valid also in the running-width
scheme. Let us finally remark that although the compensating factor
$G_{\rm FLS}$ was derived for the pure light-fermion-loop contribution
to the $W$-boson width, the respective Ward identity is satisfied
for any numerical value of $\Gamma_W$. So, in particular, one may use the 
radiatively corrected value of the $W$-width. 

It should also be noted that in order to preserve a gauge-independent 
definition of the $W$-boson mass and width beyond the leading order, 
one should use, in both the fixed- and running-width schemes, the {\em pole}
rather than {\em on-shell} $W$ mass and width, 
see e.g.\ Refs.~\cite{Passera:1998,Sirlin:1998}.

\section{The YFS exponentiation in leptonic $W$ decays}
\label{sec:YFS}

\begin{figure}[!ht]
\centering
\epsfig{file=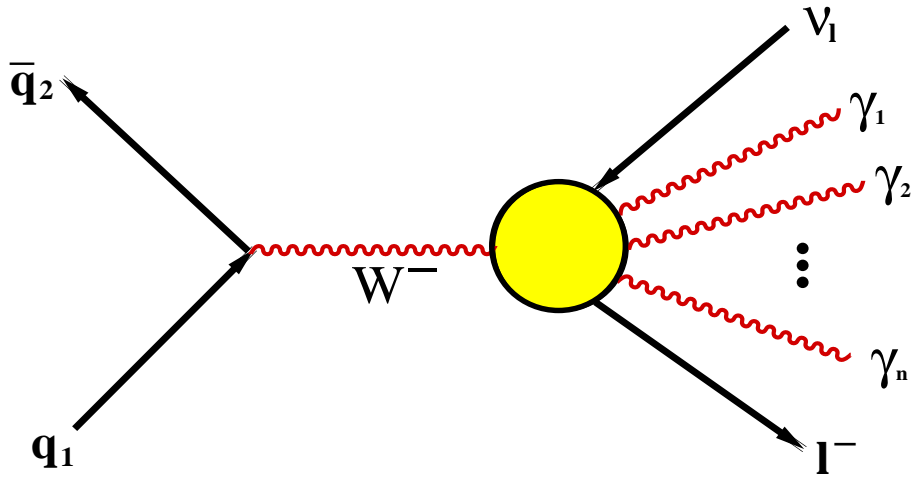,width=120mm,height=65mm}
\caption{\sf
 Production of a single $W^-$ in quark--antiquark collisions 
 with multiphoton radiation in $W$-boson decay.
}
\label{fig:WpdYFS}
\end{figure}
As was mentioned in the Introduction, the main purpose of this work is to 
provide a theoretical prediction for the multiphoton radiation in leptonic
$W$-boson decays within the YFS exclusive exponentiation scheme.
In this paper we consider the process of single-$W$ production
in hadronic collisions at the parton level, i.e.
\begin{equation}
q_1(p_1) + \bar{q}_2(p_2) \longrightarrow W^{\pm}(Q) 
 \longrightarrow l(q_l) + \nu(q_{\nu})
              + \gamma(k_1) + \ldots + \gamma(k_n),
 \:\:\: (n = 0,1,\ldots),
\label{eq:WpdYFS}
\end{equation}
depicted diagrammatically in Fig.~\ref{fig:WpdYFS}. Here we do not rely
on the small-lepton-mass approximation, i.e. the formulae below are given
for arbitrary final-state lepton masses.

The \oal\ QED YFS-exponentiated total cross section for this process 
reads
\begin{equation}
\sigma^{tot}_{\rm YFS} = \sum_{n=0}^{\infty} \int \frac{d^3q_l}{q_l^0}
  \frac{d^3q_{\nu}}{q_{\nu}^0}\rho_n^{(1)}(p_1,p_2,q_1,q_2,k_1,\ldots,k_n),
\label{eq:sigtot}
\end{equation}
where
\begin{equation}
\begin{aligned}
\rho_n^{(1)}   =  \, e^{Y(Q,q_l;k_s)}\,\frac{1}{n!} &
  \prod_{i=1}^n \frac{d^3k_i}{k_i^0} \tilde{S}(Q,q_l,k_i)\theta(k_i^0-k_s)
  \,\delta^{(4)}\left(p_1 + p_2 - q_l - q_{\nu} - \sum_{i=1}^n k_i\right)\\
  &\times  
  \left[\,\bar{\beta}_0^{(1)}(p_1,p_2,q_l,q_{\nu}) + \sum_{i=1}^n 
   \frac{\bar{\beta}_1^{(1)}(p_1,p_2,q_l,q_{\nu},k_i)}{\tilde{S}(Q,q_l,k_i)}\,
  \right];
\end{aligned}
\label{eq:rhon}
\end{equation}
here,
\begin{equation}
\tilde{S}(Q,q_l,k) = -\frac{\alpha}{4\pi^2}\left( \frac{Q}{kQ}
                                                - \frac{q_l}{kq_l}
                                           \right)^2
\label{eq:Stil}
\end{equation}
is the soft-photon radiation (eikonal) factor and
\begin{equation}
Y(Q,q_l;k_s) = 2\alpha\left[\Re B(Q,q_l;m_{\gamma}) 
                              + \tilde{B}(Q,q_l;m_{\gamma},k_s)\right]
\label{eq:Yfmf}
\end{equation}
the YFS form factor, where $B$ and $\tilde{B}$ are the virtual- and
real-photon IR YFS functions, given explicitly in Appendix~B for
arbitrary four-momenta and masses of charged particles. These IR functions are
regularized with the dummy photon mass $m_{\gamma}$, which cancels out 
in their sum. The real-photon function $\tilde{B}$ depends also on the 
soft-photon energy cut-off $k_s \ll E_{CM}$, which means that it was 
integrated analytically over the photons with energies $E_{\gamma}\leq k_s$.
The photons with energies $E_{\gamma} > k_s$ are generated exclusively 
with the help of Monte Carlo techniques. 
The soft cut-off $k_s$ is a dummy parameter,
i.e. the resulting cross section does not depend on it, which can be checked
both analytically (e.g. by differentiating Eq.~(\ref{eq:rhon}) over $k_s$)
and numerically (by evaluating the cross section for different values of
$k_s$). One of the advantages of exponentiation is that $k_s$ can be put
arbitrarily low without causing any part of the cross section to become
negative -- in contrast to fixed-order calculations. In Eq.~(\ref{eq:rhon}),
$\bar{\beta}_0^{(1)}$ and $\bar{\beta}_1^{(1)}$ are the YFS non-IR functions, 
calculated perturbatively through \oal. We present them below in the
centre-of-mass (CM) frame of the incoming quarks, i.e. the rest frame of $W$,
with the $+z$ axis pointing in the quark $q_1$ direction. 

The function $\bar{\beta}_0^{(1)}$ is given by
\begin{equation}
\bar{\beta}_0^{(1)}(p_1,p_2,q_l,q_{\nu}) = 
\bar{\beta}_0^{(0)}(p_1,p_2,q_l,q_{\nu})
\left[\, 1 + \delta^{(1)}(Q,q_l,q_{\nu})\,\right],
\label{eq:beta01}
\end{equation}
where $\bar{\beta}_0^{(0)}$ is related to the Born-level cross section through
\begin{equation}
\frac{1}{2}\bar{\beta}_0^{(0)} = \frac{1}{\sqrt{\lambda(1,m_l/M,m_{\nu}/M)}}\,
  \frac{d\sigma_0}{d\Omega_l} =
  \frac{1}{16s\,(2\pi)^2}\, \frac{1}{12}
  \sum \left|{\cal M}^{(0)}\right|^2,
\label{eq:beta00}
\end{equation}
with $s=(q_1+q_2)^2=Q^2$ and 
 $\lambda(x,y,z) = x^2 + y^2 + z^2 - 2xy - 2xz - 2 yz$. The factor
$\frac{1}{12}=\frac{1}{4}\cdot\frac{1}{3}$ corresponds to averaging over the 
initial-state quark spins and colours 
(the colour contents has been extracted explicitly), 
and the sum $\sum$ runs over all the initial- and final-state spin indices.
In  Eq.~(\ref{eq:beta01}), the correction
\begin{equation}
\delta^{(1)}(Q,q_l,q_{\nu}) = \delta^v_{\rm EW}(Q,q_l,q_{\nu};m_{\gamma}) 
                            - 2\alpha\Re B(Q,q_l;m_{\gamma})
\label{eq:delta1}
\end{equation}
is the 1st order non-IR correction to the $\bar{\beta}_0$ function, where
$\delta^v_{\rm EW}$ is the \oal\ EW virtual correction.
Since in this paper we limit ourselves to the QED-like corrections,
from Eqs.~(\ref{eq:vir1}) and (\ref{eq:ReBtzsm}) we have
\begin{equation}
 \delta_{\rm QED}^{(1)}(Q,q_l) = 
 \frac{\alpha}{\pi}\left(\ln\frac{M}{m_l} + \frac{1}{2}\right).
\label{eq:delQED}
\end{equation}
Although Eqs.~(\ref{eq:vir1}) and (\ref{eq:ReBtzsm}) were obtained 
in the small-lepton-mass approximation, $m_l \ll M$, we have checked
that the above formula remains true for arbitrary lepton mass $m_l < M$
(of course, under the assumption that $\delta_{\rm QED}^v$ contains only
the IR- and mass-singular terms).

The function $\bar{\beta}_1^{(1)}$ is the YFS non-IR function corresponding
to the single-real-hard photon radiation. It is related to  
differential cross sections through
\begin{equation}
\frac{1}{2}\bar{\beta}_1^{(1)}(p_1,p_2,q_l,q_{\nu},k) 
 = \frac{1}{\sqrt{\Lambda(k)}}\,\frac{d\sigma_1}{d\Omega_lk^0dk^0d\Omega_k}
   - \tilde{S}(Q,q_l,k)\,  \frac{1}{\sqrt{\lambda}}\,
    \frac{d\sigma_0}{d\Omega_l},
\label{eq:beta11}
\end{equation}
where 
\begin{equation}
\frac{d\sigma_1}{d\Omega_lk^0dk^0d\Omega_k} = 
     \frac{\sqrt{\Lambda(k)}}{32s\,(2\pi)^5}
     \, \frac{1}{12} \sum \left|{\cal M}^{(1)}\right|^2,
\label{eq:dsig1}
\end{equation}
with 
\begin{equation}
\sqrt{\Lambda(k)} = \frac{2\,|\vec{q}_l|^2}{|\vec{q}_l|(M - k^0) 
                + q_l^0|\vec{k}|\cos\theta_{lk}}
\label{eq:LaPS}
\end{equation}
the phase-space factor (coming from the phase-space integration 
eliminating the energy-momentum conservation $\delta^{(4)}$-function
for single-photon radiation), 
where $\theta_{lk} = \angle(\vec{q}_l,\vec{k})$; in the soft-photon limit
$\Lambda(k\rightarrow 0) \rightarrow \lambda$. 
The sum $\sum$ in Eq.~(\ref{eq:dsig1}) again runs over the initial- 
and final-state spin indices, this time inluding also those of the
radiative photon.
Thus, we finally have
\begin{equation}
\bar{\beta}_1^{(1)}(p_1,p_2,q_l,q_{\nu},k) 
 = \frac{1}{16s\,(2\pi)^5} 
   \, \frac{1}{12} \sum \left|{\cal M}^{(1)}\right|^2
    - \tilde{S}(Q,q_l,k)\bar{\beta}_0^{(0)}(p_1,p_2,q_l,q_{\nu}).
\label{eq:beta11f}
\end{equation}

There are several advantages in using the matrix elements 
of Section~\ref{sec:sa}.
Firstly, the respective spin amplitudes are derived without the assumption of
the energy-momentum conservation. Therefore, they can be used directly in 
evaluations of the above YFS $\bar{\beta}$-functions over the multiphoton
phase space, without the need to resort to any ``reduction 
procedure'', which reduces the multiphoton phase space to the $0$-photon
phase space for $\bar{\beta}_0$ and the $1$-photon phase space for 
$\bar{\beta}_1$, see e.g.~\cite{yfs:1961,yfs2:1990}. 
Secondly, since the spin amplitudes are obtained for massive fermions,
there is no need to use any phase-space slicing or subtraction methods
in order to separate mass singularities~\cite{Dittmaier:2001ay}.
Using spin amplitudes instead of explicit analytical formulae for the squared
matrix elements may also be useful for some dedicated studies, such as
investigation of various $W$-polarization contributions,
``new physics'' searches (spin amplitudes can be easily modified to 
include some ``new physics'' components), etc.   
And, which is important in practice, the numerical evaluation of the matrix 
elements based on the above spin amplitudes is fast in terms of CPU time.

In computing the matrix element 
$\sum\left|{\cal M}^{(1)}\right|^2$ we observed a loss of numerical
precision $\sim$~\Order{0.1\%} when the angle between the radiative photon 
and the electron (positron), $\theta_{e\gamma}$, was $\sim$~\Order{10^{-6}}. 
It turned out that most of this precision loss was coming from huge numerical 
cancellations between the terms in the universal eikonal factor 
of Eq.~(\ref{eq:saWd1}) (the factor in front of the first $S$-function).
We improved this by correcting the above matrix element according to
\begin{equation}
 \sum \left|{\cal M}^{(1)}\right|^2 \longrightarrow 
 \sum \left|{\cal M}^{(1)}\right|^2 + \delta_{coll},
\label{eq:M1cor}
\end{equation}
where
\begin{equation}
\delta_{coll} = \left[\,16\pi^3\tilde{S}(Q,q_l,k) 
 - e^2 
 \sum_{\kappa}\left|\frac{q_l\cdot\epsilon_{\gamma}^{\ast}(\kappa)}{k\cdot q_l}
   -\frac{Q\cdot\epsilon_{\gamma}^{\ast}(\kappa)}{k\cdot Q}\right|^2\, \right]
 \sum \left|{\cal M}^{(0)}\right|^2.
\label{eq:delcol}
\end{equation}
Algebraically, the two terms in the square brackets
are identical. Numerically, however, they can differ for ultra-collinear
photon radiation, owing to huge cancellations in the second term leading 
to a loss of numerical precision. Therefore, this correction effectively
replaces the numerically unstable part of the matrix element 
$\sum\left|{\cal M}^{(1)}\right|^2$ corresponding to the second term
in Eq.~(\ref{eq:delcol}) with the numerically safe one corresponding
to the first term, obtained directly from the particles four-momenta. 
We have checked that the above modification is sufficient for the numerical 
precision of \Order{10^{-4}} for $\theta_{e\gamma}\lesssim 10^{-6}$
and of \Order{10^{-8}} for the total cross section. 
By looking at Eq.~(\ref{eq:beta11f}) one can notice that the part of
the matrix element $\sum\left|{\cal M}^{(1)}\right|^2$ that is proportional
to the soft-photon factor $\tilde{S}$ exactly cancels in the calculation
of the $\bar{\beta}_1^{(1)}$ function. We could, therefore, perform this
cancellation algebraically and thus avoid the above numerical problems. 
We, however, keep this term in $\sum\left|{\cal M}^{(1)}\right|^2$ because
apart from the YFS exponentiation we want to have in our program also
the non-exponentiated, fixed-order \oal\ calculation. 
Since it is now calculated in the same way as the second term in 
Eq.~(\ref{eq:beta11f}),  it exactly cancels numerically
in the evaluation of the $\bar{\beta}_1^{(1)}$ function. 

This completes our description of the cross section for 
process~(\ref{eq:WpdYFS}) with the \oal\ QED YFS exponentiation.
In order to compute this cross section and generate events, we have
developed an appropriate MC algorithm, which will be described in detail
elsewhere~\cite{WINHAC:2002}. We will complete this paper by presenting
some results of numerical tests of the corresponding MC program, called
WINHAC.  

\section{Numerical results}
\label{sec:results}

We performed several numerical tests of the MC event generator
WINHAC, which implements the calculations presented above.
Here we discuss some of the results. We considered the following
process:
\begin{equation}
d + \bar{u} \longrightarrow W^-  \longrightarrow l + \bar{\nu_l},
\label{eq:respro}
\end{equation}
where $l=e,\,\mu,\,\tau$. We have checked that the results remain unchanged
when we switch to the corresponding process of $W^+$ production and decay.
Our MC calculations were done using the $G_{\mu}$ scheme and the fixed-width
scheme. All the results below, unless stated
otherwise, have been obtained  for the following input parameters:
\begin{equation}
\begin{aligned}
\,& m_d = 3\times 10^{-3}\,{\rm GeV},\:\:\:\:\:\: m_u = 6\times 10^{-3},
  \:\:\:\:\:\: V_{ud} = 1, \:\:\:\:\:\: m_{\nu_l} = 0,\\
\,& m_e = 0.511\times 10^{-3}\,{\rm GeV},\:\:\:\:\:\:
  m_{\mu} = 0.10565836\,{\rm GeV},\:\:\:\:\:\:
  m_{\tau} = 1.77703\,{\rm GeV}, \\
\,& M_W = 80.423\,{\rm GeV},\:\:\:\:\:\:  M_Z = 91.1882\,{\rm GeV}\\
\,& s_W^2 = 1 -\frac{M_W^2}{M_Z^2}, \:\:\:\:\:\: 
\Gamma_W = \frac{3G_{\mu}M_W^3}{2\sqrt{2}\pi}\left(1 
           + \frac{2\alpha_s}{3\pi}\right), \\
\,& \alpha = 137.03599976,\:\:\:\:\:\:
G_{\mu} = 1.16639 \times 10^{-5}\,{\rm GeV}^{-2}, \:\:\:\:\:\:
\alpha_s = 0.1185, \\
\,& E_{CM} = \sqrt{s} = M_W. \\
\end{aligned}
\label{eq:input}
\end{equation}

\subsection{General tests}
\label{ssec:tests}

We have performed several numerical tests of the MC event generator
WINHAC aimed at checking the correctness of the implemented matrix 
elements as well as the corresponding MC algorithm. 

In order to cross-check the matrix elements presented here,
we implemented in our MC program the matrix elements
of Ref.~\cite{Berends:1985}, which in the following we shall call B\&K.
These latter matrix elements were obtained in the small-lepton-mass
approximation $m_l\ll M_W$; their precision therefore is of
\Order{m_l^2/M_W^2}, which for electrons gives \Order{10^{-10}}.
Since our spin amplitudes are obtained for massive fermions, we performed
the comparisons of these matrix elements for electronic
$W$-boson decays. We did this by taking the difference between
the corresponding MC weights on an event-by-event basis and calculating
the average of this difference over the whole MC sample. For both
the Born-level and \oal\ matrix elements, we reached an agreement
at the level of $\sim 10^{-8}$.  

Then, we performed several tests to check the MC algorithm of the program
WINHAC. An important test of the algorithm for MC integration and event 
generation according to Eq.~(\ref{eq:sigtot}) is to reproduce
fixed-order calculations. The strict Born-level cross section can be obtained
from Eq.~(\ref{eq:sigtot}) by truncating the perturbation series in $\alpha$
at the lowest-order term, which amounts to
\begin{equation} 
\sigma_0^{tot} = \int\frac{d^3q_l}{q_l^0}\frac{d^3q_{\nu}}{q_{\nu}^0}\,
                 \rho_0^{(0)}\,e^{-Y}\,.
\label{eq:bornmc}
\end{equation}
Within the multiphoton MC algorithm, this means calculating an appropriate
weight if the photon number $n=0$ and setting it to zero if $n>0$.
The Born-level total cross section can be easily calculated analytically.
In the small-fermion-mass approximation and in  the fixed-width scheme 
it reads
\begin{equation}
\sigma_0^{tot} = \frac{\alpha_{G_{\mu}}^2\pi |V_{q_1q_2}|^2}{36s_W^4}\,
                 \frac{s}{(s-M_W^2)^2 + M_W^2\Gamma_W^2}\,.
\label{eq:bornan}
\end{equation}
\begin{table}[!ht]
\centering
\begin{tabular}{||l|c|c|c||}
\hline\hline
\raisebox{-1.5ex}[0cm][0cm]{Calculation} & 
\multicolumn{3}{|c||}{$\sigma_0^{tot}$ [nb]} \\
\cline{2-4}
            & $e$           & $\mu$         & $\tau$ \\
\hline\hline
Analytical & 
   $ 8.8872\:\:\:\:\:\;$ & $8.8872\:\:\:\:\:\;$ & $8.8872\:\:\:\:\:\;$\\
\hline
WINHAC      & $8.8869\,(2)$ & $8.8873\,(2)$ & $8.8808\,(2)$ \\
\hline\hline
\end{tabular}
\caption{\small\sf
  The results for the total Born-level cross section from the
  MC program WINHAC compared with the analytical calculation in the 
  small-fermion-mass approximation.
  The numbers in parentheses are statistical errors 
  for the last digits.
}
\label{tab:born}
\end{table}
In Table~\ref{tab:born} we compare the results for the total Born
cross section for $e$, $\mu$ and $\tau$ in the final state, calculated
with the MC program WINHAC with those obtained from the analytical
formula of  Eq.~(\ref{eq:bornan}). We see a very good agreement 
between these two calculations for $e$ and $\mu$. For $\tau$ they
differ by $\sim 0.1\%$, which can be explained by the 
$\tau$-mass effects (they are not negligible as in the case of
$e$ and $\mu$).

In a similar way, the first-order cross section can be obtained from
Eq.~(\ref{eq:sigtot}) by truncating the perturbative series at
\oal\ beyond the Born level, i.e.
\begin{equation} 
\begin{aligned}
\sigma_1^{tot} &= \int\frac{d^3q_l}{q_l^0}\frac{d^3q_{\nu}}{q_{\nu}^0}\,
                 \delta^{(4)}(p_1+p_2-q_l-q_{\nu})\,
   \bar{\beta}_0^{(0)}\left[\,1 + \delta_{\rm QED}^{(1)} + Y\,\right] \\
  & + \int\frac{d^3q_l}{q_l^0}\frac{d^3q_{\nu}}{q_{\nu}^0}\frac{d^3k}{k^0}\,
      \delta^{(4)}(p_1+p_2-q_l-q_{\nu}-k)
  \left[\,\bar{\beta}_1^{(1)} + \tilde{S}\bar{\beta}_0^{(0)}\,\right]
  \theta(k^0-k_s)\,,
\end{aligned}
\label{eq:firstmc}
\end{equation}
where the first term on the r.h.s. corresponds to the Born plus virtual
and real-soft-photon contribution, and the second term to 
the real-hard-photon contribution. 
In practice, this means that the first term is evaluated
within the multiphoton algorithm only for $n=0$, the second only for $n=1$,
otherwise the appropriate MC weights are set to zero. 
\begin{table}[!ht]
\centering
\begin{tabular}{||l|c|c|c||}
\hline\hline
\raisebox{-1.5ex}[0cm][0cm]{Calculation} & 
\multicolumn{3}{|c||}{$\delta_1 = \sigma_1^{tot}/\sigma_0^{tot}-1$} \\
\cline{2-4}
            & $e$           & $\mu$         & $\tau$ \\
\hline\hline
WINHAC      & $-1.5\,(3)\times 10^{-4}$ & $-2.2\,(3)\times 10^{-4}$ 
            & $-0.3\,(2)\times 10^{-4}$ \\
\hline\hline
\end{tabular}
\caption{\small\sf
  The results for the \oal\ QED-like correction to the total cross section 
  from the MC program WINHAC.
  The numbers in parentheses are statistical errors 
  for the last digits.
}
\label{tab:delt1}
\end{table}
In Table~\ref{tab:delt1} we show the results from the program WINHAC
for the pure \oal\ QED-like correction to the total cross section. 
As can be seen, the results for $e$ and $\mu$ are in good agreement with 
the numerical value of total QED-like correction to the $W$-boson 
width as given in Eq.~(\ref{eq:QED1}). For $\tau$ we observe the difference
of $\sim 1.5\times 10^{-4}$, which again can be explained by the $\tau$-mass
effects.

\begin{table}[!ht]
\centering
\begin{tabular}{||c|r|r|r|r||}
\hline\hline
\raisebox{-1.5ex}[0cm][0cm]{$k_0$} & 
\multicolumn{2}{|c|}{$e$} & \multicolumn{2}{|c||}{$\mu$}\\
\cline{2-5}
            & WINHAC & B\&K &  WINHAC & B\&K \\
\hline
$0.01$ & $19.69\,(3)$ & $19.7$ & $10.11\,(2)$ & $10.1$ \\
$0.05$ & $11.61\,(2)$ & $11.6$ & $ 5.92\,(1)$ & $ 5.9$ \\
$0.10$ & $ 8.31\,(2)$ & $ 8.3$ & $ 4.22\,(1)$ & $ 4.2$ \\
$0.15$ & $ 6.47\,(2)$ & $ 6.5$ & $ 3.27\,(1)$ & $ 3.3$ \\
$0.20$ & $ 5.23\,(1)$ & $ 5.2$ & $ 2.63\,(1)$ & $ 2.6$ \\
$0.30$ & $ 3.61\,(1)$ & $ 3.6$ & $ 1.80\,(1)$ & $ 1.8$ \\
$0.40$ & $ 2.57\,(1)$ & $ 2.6$ & $ 1.27\,(1)$ & $ 1.3$ \\
$0.50$ & $ 1.84\,(1)$ & $ 1.8$ & $ 0.91\,(1)$ & $ 0.9$ \\
$0.60$ & $ 1.29\,(1)$ & $ 1.3$ & $ 0.63\,(1)$ & $ 0.6$ \\
$0.70$ & $ 0.86\,(1)$ & $ 0.9$ & $ 0.42\,(1)$ & $ 0.4$ \\
$0.80$ & $ 0.52\,(1)$ & $ 0.5$ & $ 0.25\,(1)$ & $ 0.2$ \\
$0.90$ & $ 0.24\,(1)$ & $ 0.2$ & $ 0.11\,(1)$ & $ 0.1$ \\
\hline\hline
\end{tabular}
\caption{\small\sf
  The fraction of events (in \%) with a photon energy greater than $k_0$
  at \oal\ from the MC program WINHAC and from the MC program of 
  Berends \& Kleiss~\cite{Berends:1985} (denoted as B\&K) for $E_{CM}=90\,$GeV.
  The numbers in parentheses are statistical errors 
  for the last digits.
}
\label{tab:phosp}
\end{table}
In Table~\ref{tab:phosp} we compare the results for the \oal\ 
hard-photon correction as a function of the lower photon-energy 
cut-off $k_0$, i.e.
\begin{equation}
\delta_1^h(k_0) = \frac{1}{\sigma_1^{tot}} \int_{k_0} dE_{\gamma}\,
                  \frac{d\sigma_1}{E_{\gamma}} \; \times 100\%\,,
\label{eq:phosp}
\end{equation}
for the centre-of-mass energy $E_{CM}=90\,$GeV, 
obtained from the program WINHAC and from the
B\&K MC program~\cite{Berends:1985}.  
The results of these two programs agree very well within the statistical
errors.

As the above fixed-order results from WINHAC have been obtained in the
framework of the YFS-type multiphoton algorithm, they make us
strongly confident in the correctness of the corresponding MC algorithm.   

\begin{table}[!ht]
\centering
\begin{tabular}{||c|c|c|c||}
\hline\hline
\raisebox{-1.5ex}[0cm][0cm]{Calculation} & 
\multicolumn{3}{|c||}{$\sigma^{tot}\,$[nb]} \\
\cline{2-4}
     & $e$           & $\mu$         & $\tau$ \\
\hline\hline
Fixed \oal-level   & $8.88564\,(14)$ & $8.88539\,(12)$ & $8.88047\,(10)$ \\
YFS exponentiation & $8.88390\:\:\:(6)$ & $8.88443\:\:\:(9)$ &
                   $8.87859\:\:\:(9)$ \\
\hline
$\delta_{exp}=(\sigma_{\rm YFS}^{tot} - \sigma_1^{tot})/\sigma_0^{tot}$ 
  & $-2.0\,(1)\times 10^{-4}$ 
  & $-1.1\,(1)\times 10^{-4}$ 
  & $-2.1\,(0)\times 10^{-4}$ \\
\hline\hline
\end{tabular}
\caption{\small\sf
  The results for the fixed-\oal\ and the YFS-exponentiated total cross 
  section  from the MC program WINHAC.
  The numbers in parentheses are statistical errors 
  for the last digits.
}
\label{tab:totyfs}
\end{table}
In Table~\ref{tab:totyfs} we give the results for the total
cross section at the fixed \oal-level and including the YFS
exponentiation as given in Eq.~(\ref{eq:sigtot}). 
The YFS-exponentiation corrections beyond \oal\ are $\sim 10^{-4}$,
i.e. of the expected size of higher-order corrections. 

\subsection{Distributions}
\label{ssec:distr}

\begin{figure}[!ht]
\centering
 
 
\setlength{\unitlength}{0.1mm}
\begin{picture}(1600,1480)
\put( 450,1420){\makebox(0,0)[cb]{\bf BARE} }
\put(1250,1420){\makebox(0,0)[cb]{\bf CALO} }
 
\put(  0,700){\makebox(0,0)[lb]{
\epsfig{file=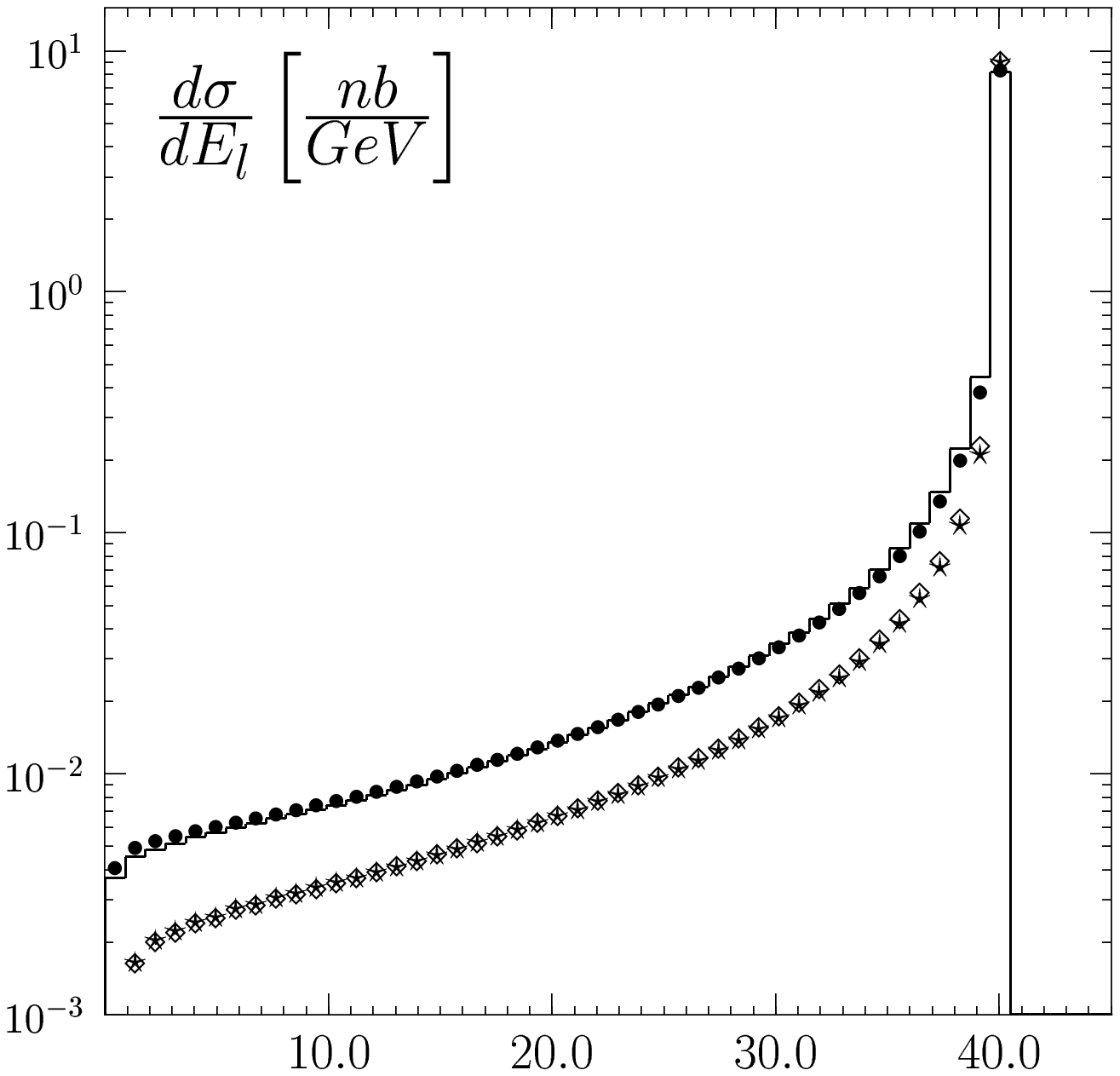                ,width=80mm,height=70mm}
}}
  
\put(  0,  0){\makebox(0,0)[lb]{
\epsfig{file=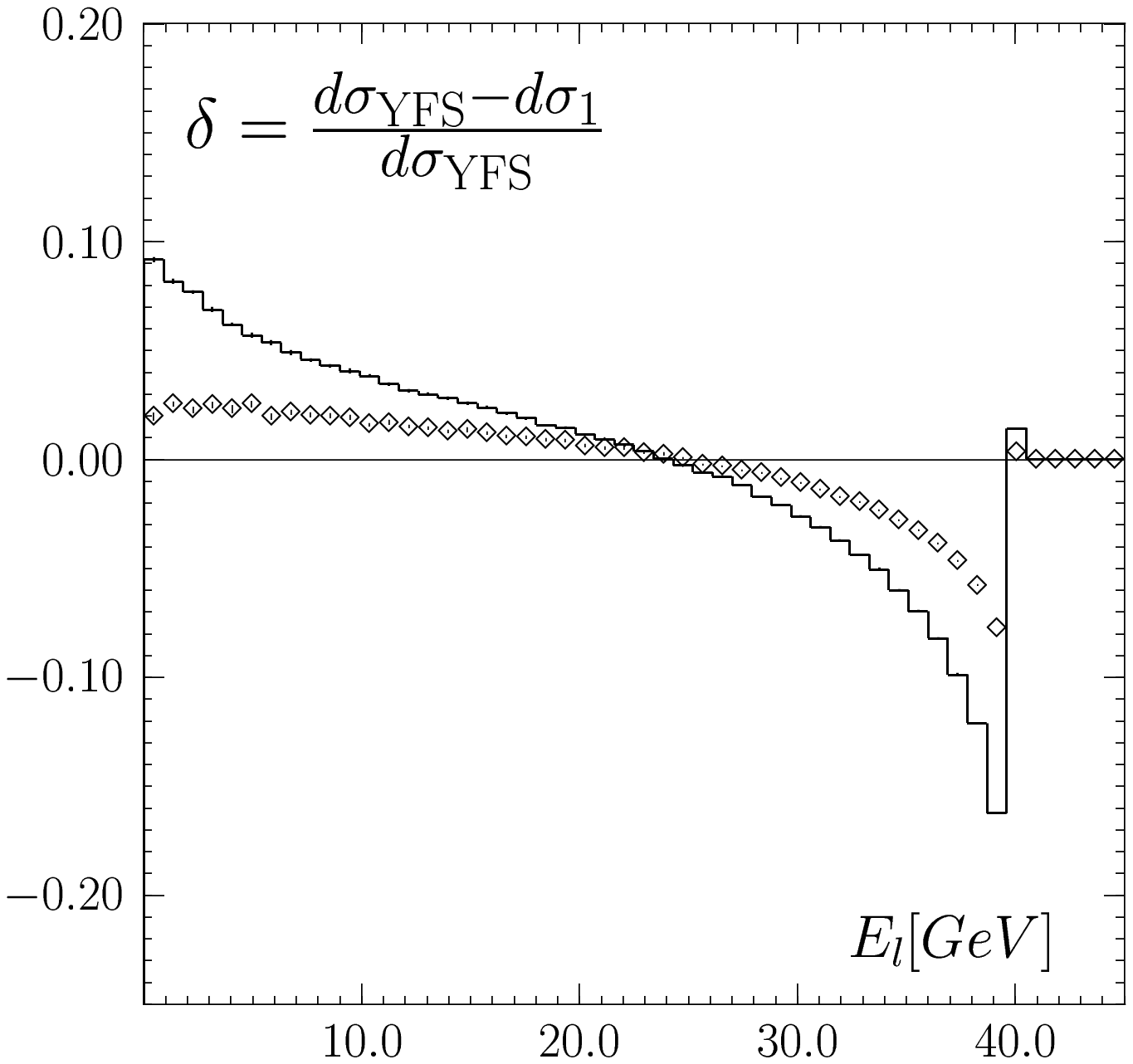            ,width=80mm,height=70mm}
}}
  
\put(800,700){\makebox(0,0)[lb]{
\epsfig{file=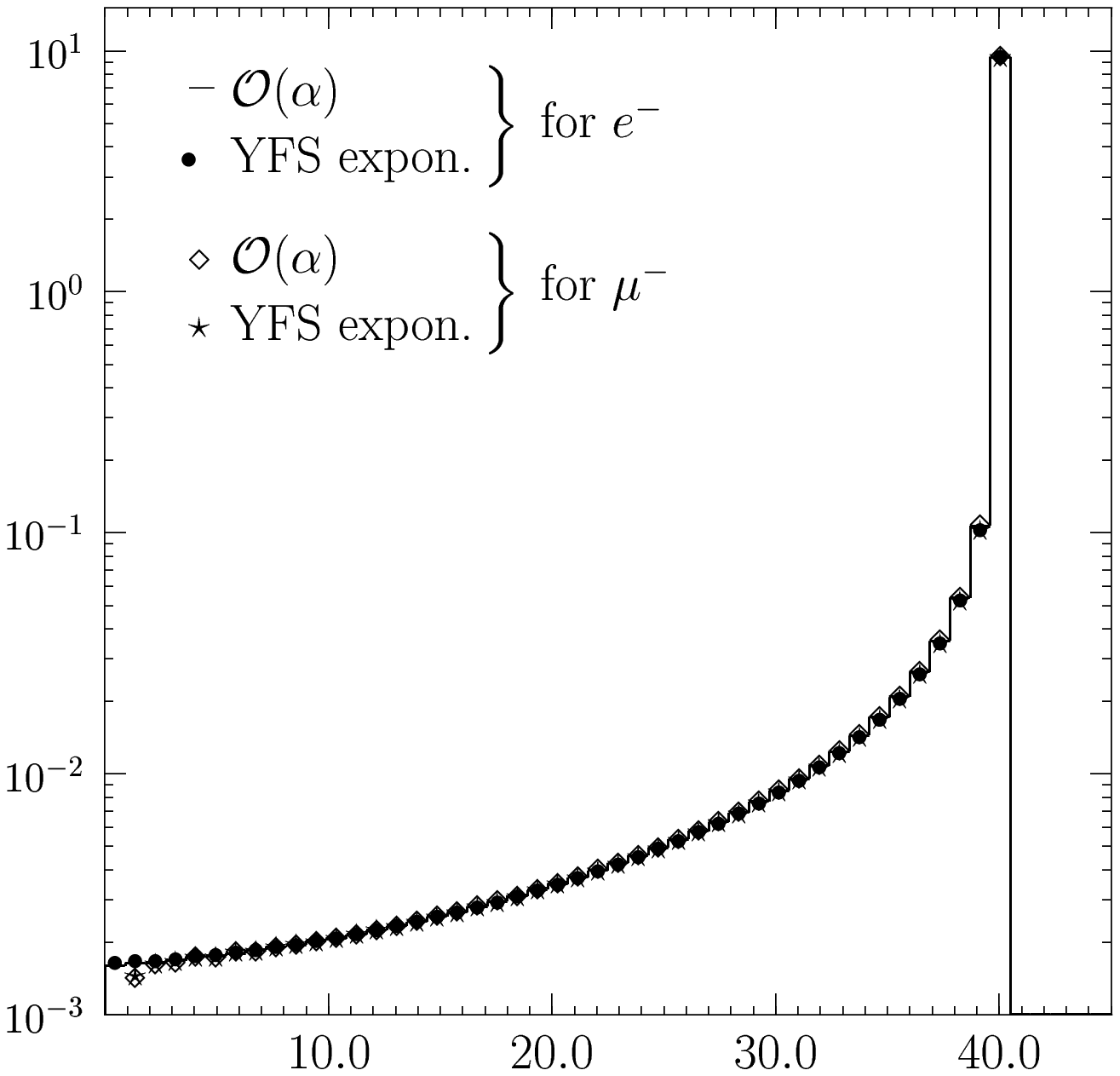                ,width=80mm,height=70mm}
}}
  
\put(800,  0){\makebox(0,0)[lb]{
\epsfig{file=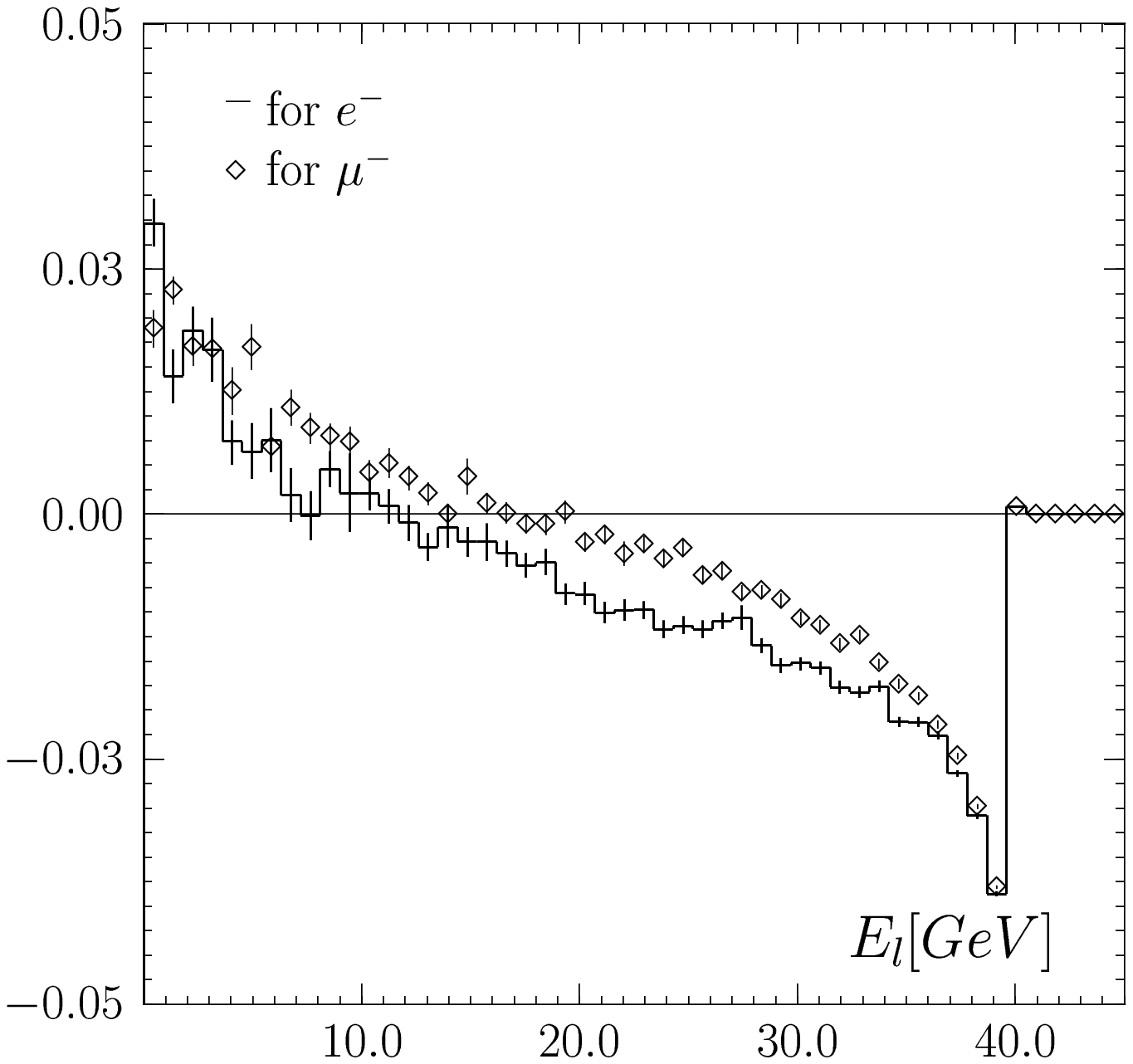            ,width=80mm,height=70mm}
}}
\end{picture}
\caption{\small\sf
 Distributions of the charged-lepton energy                           
 for BARE and CALO acceptances.                                  
}
\label{fig:El}        
\end{figure}

\begin{figure}[!ht]
\centering
 
 
\setlength{\unitlength}{0.1mm}
\begin{picture}(1600,1480)
\put( 450,1420){\makebox(0,0)[cb]{\bf BARE} }
\put(1250,1420){\makebox(0,0)[cb]{\bf CALO} }
 
\put(  0,700){\makebox(0,0)[lb]{
\epsfig{file=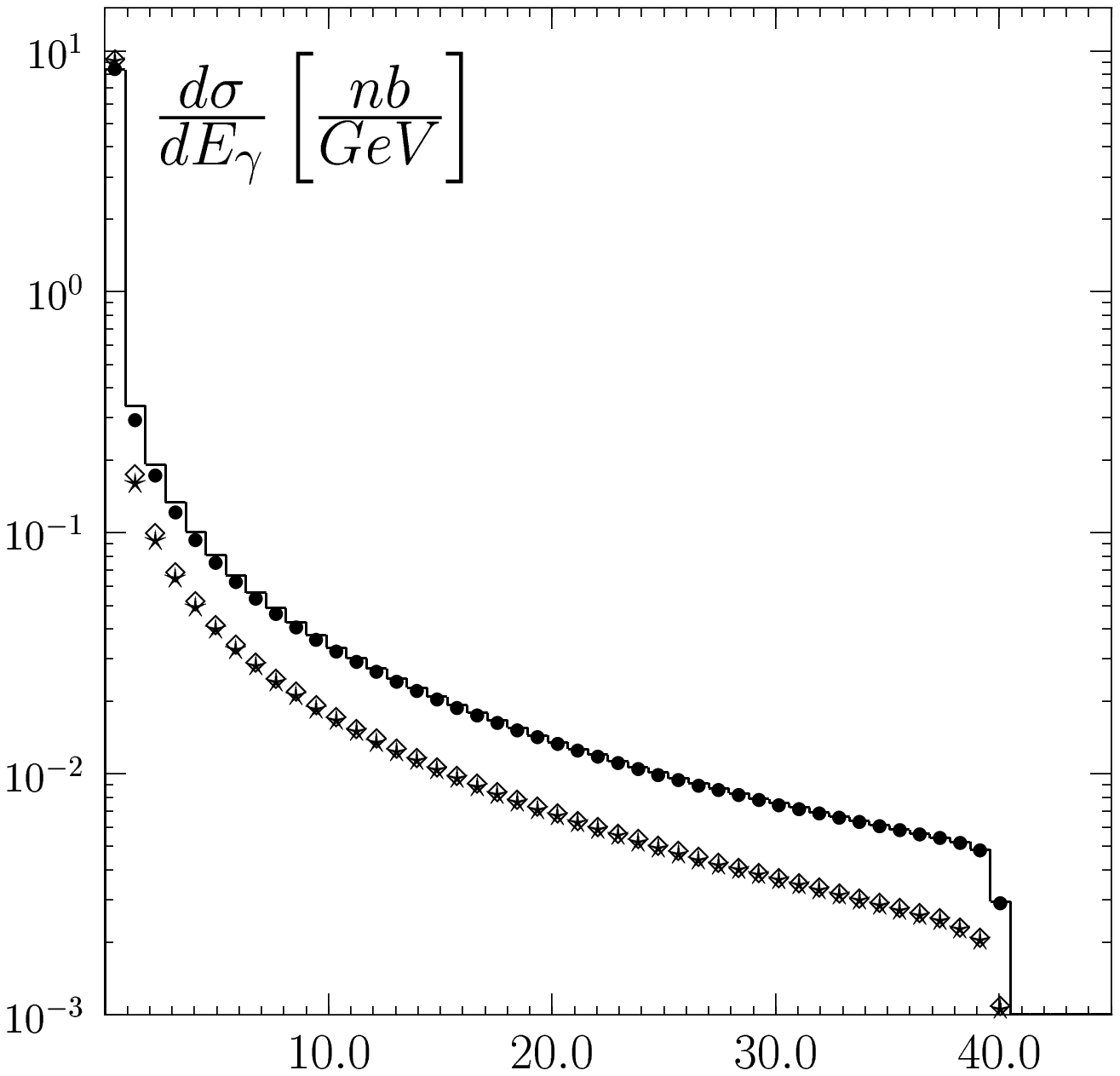                ,width=80mm,height=70mm}
}}
  
\put(  0,  0){\makebox(0,0)[lb]{
\epsfig{file=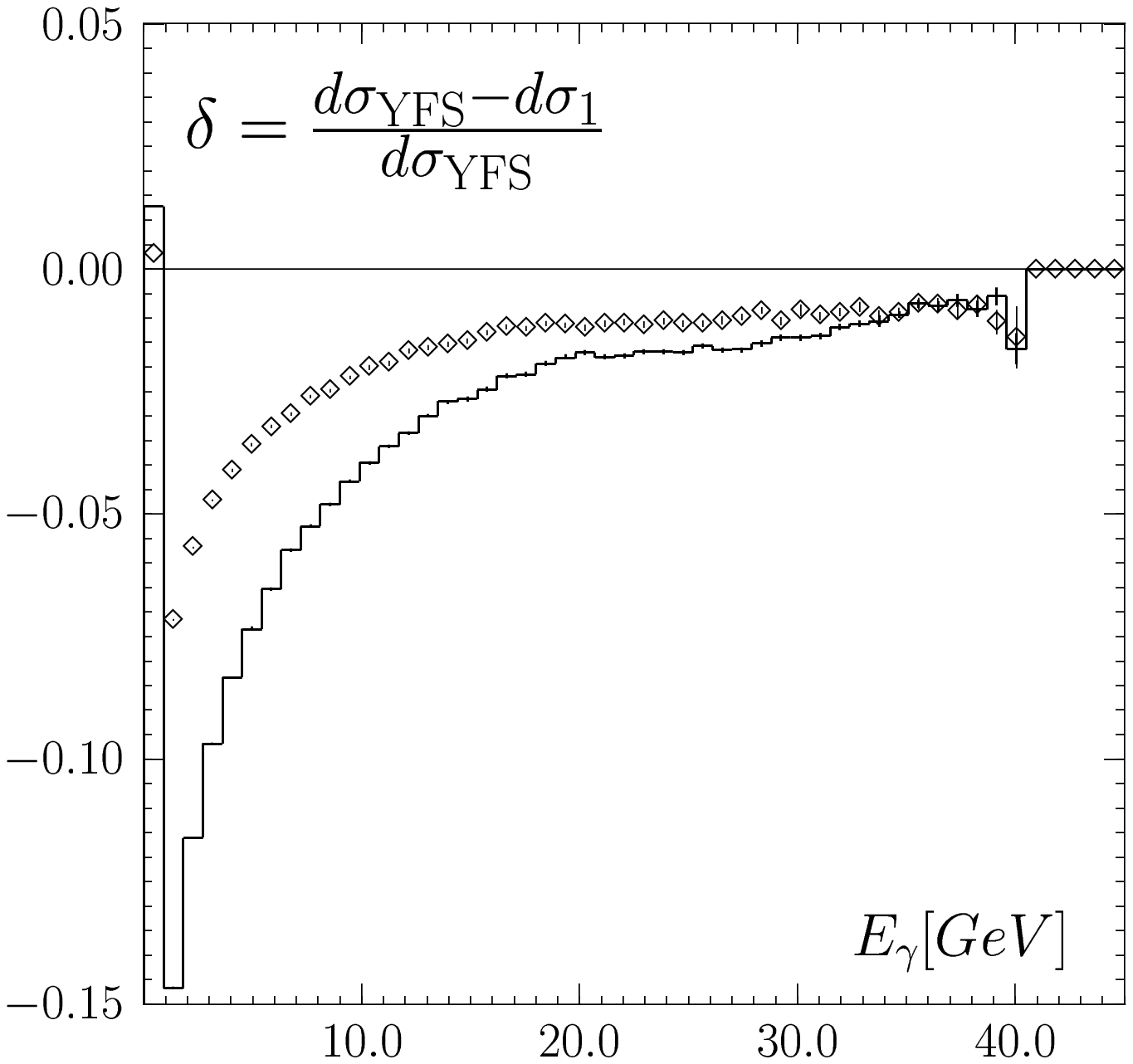            ,width=80mm,height=70mm}
}}
  
\put(800,700){\makebox(0,0)[lb]{
\epsfig{file=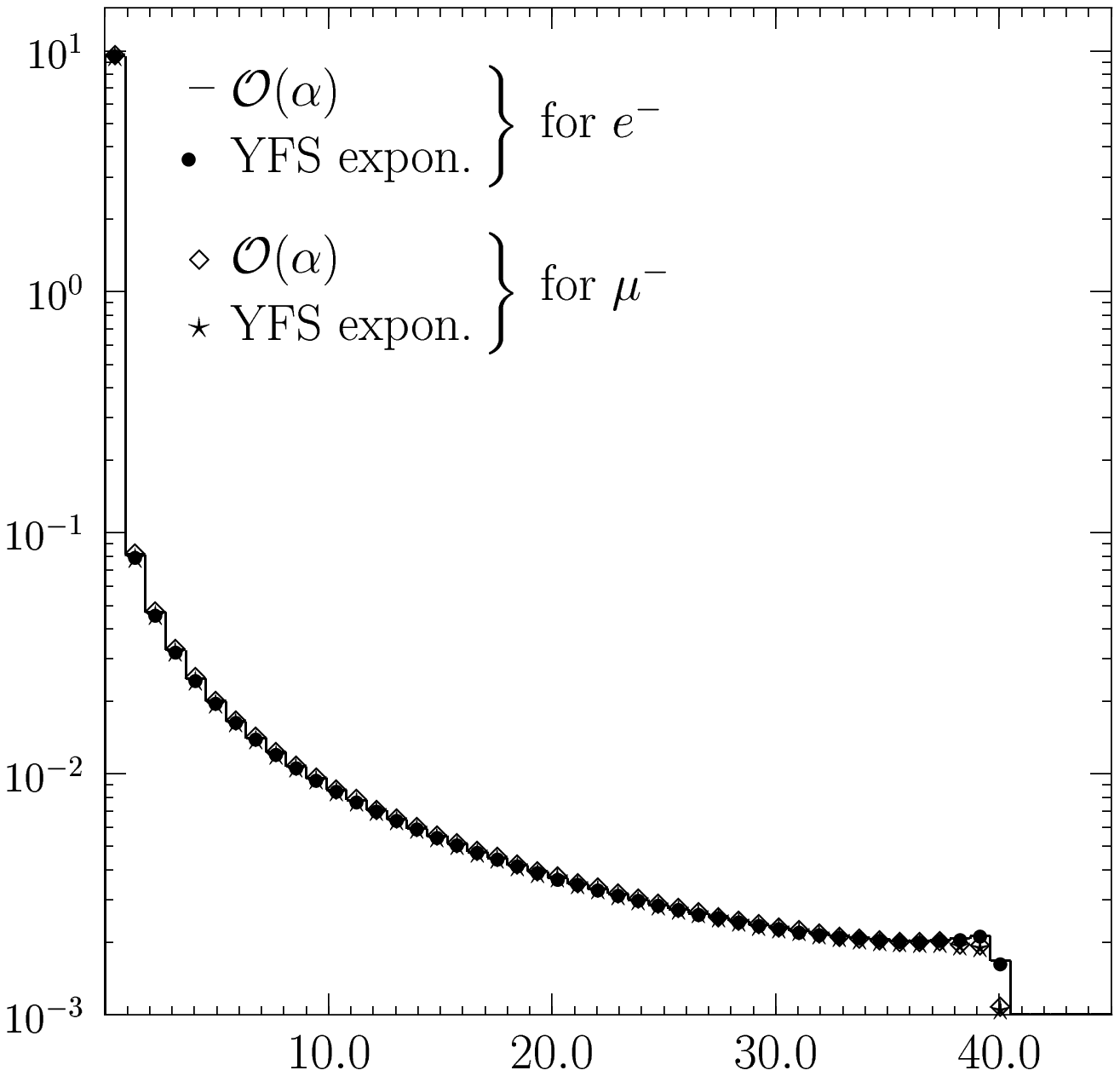                ,width=80mm,height=70mm}
}}
  
\put(800,  0){\makebox(0,0)[lb]{
\epsfig{file=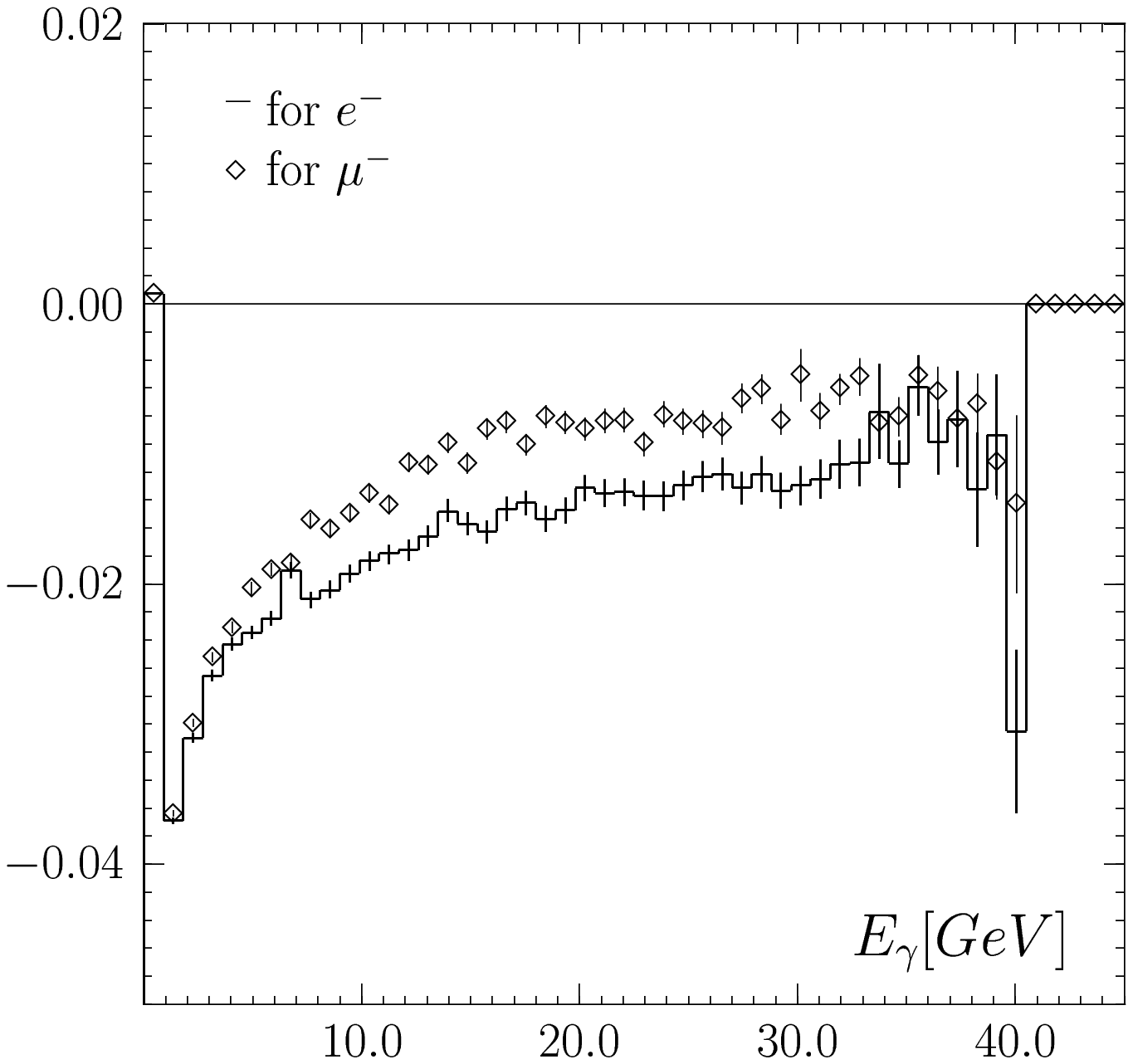            ,width=80mm,height=70mm}
}}
\end{picture}
\caption{\small\sf
 Distributions of the hardest-photon energy                     
 for BARE and CALO acceptances.                                  
}
\label{fig:Eg}        
\end{figure}

\begin{figure}[!ht]
\centering
 
 
\setlength{\unitlength}{0.1mm}
\begin{picture}(1600,1480)
\put( 450,1420){\makebox(0,0)[cb]{\bf BARE} }
\put(1250,1420){\makebox(0,0)[cb]{\bf CALO} }
 
\put(  0,700){\makebox(0,0)[lb]{
\epsfig{file=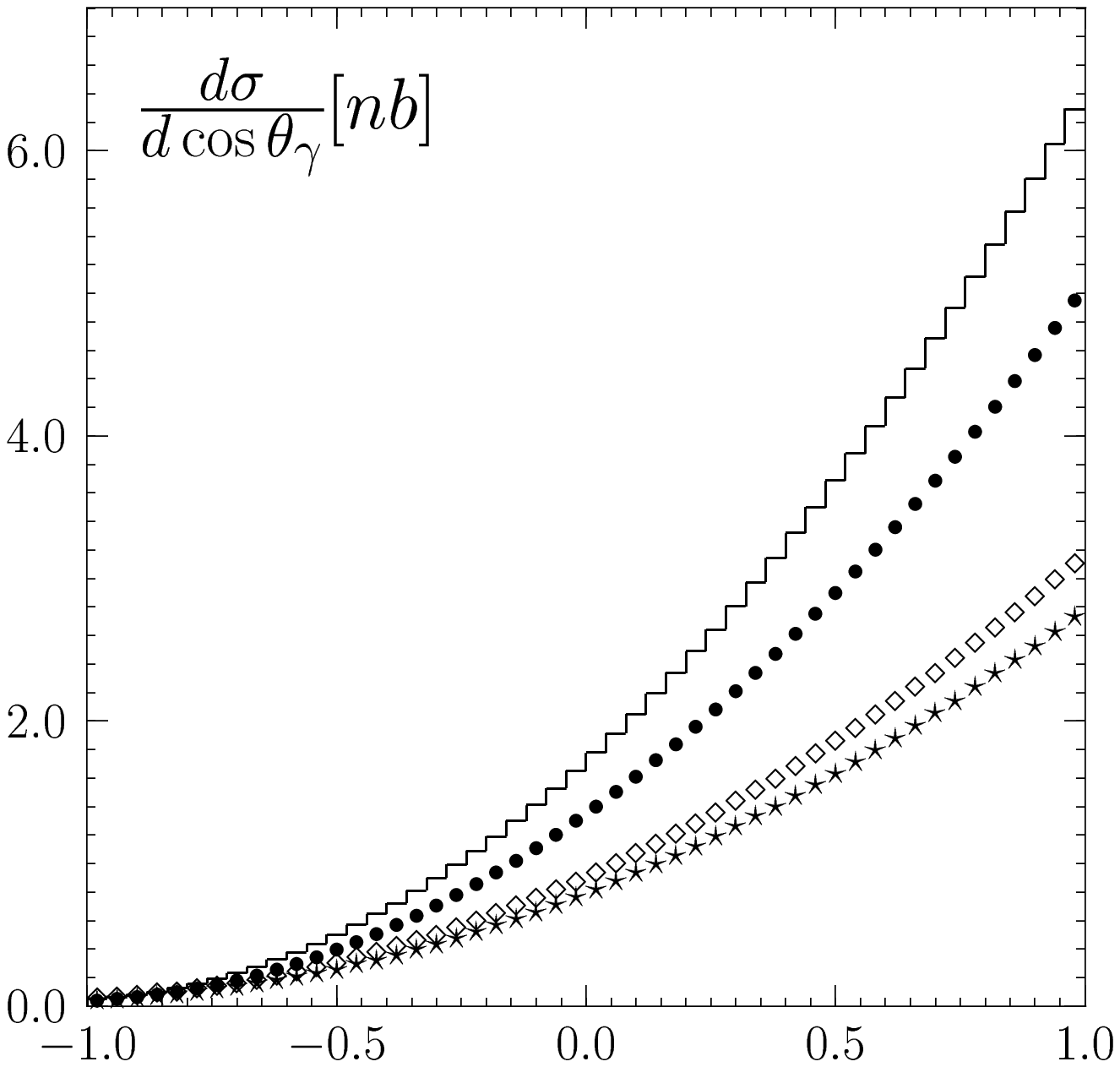                ,width=80mm,height=70mm}
}}
  
\put(  0,  0){\makebox(0,0)[lb]{
\epsfig{file=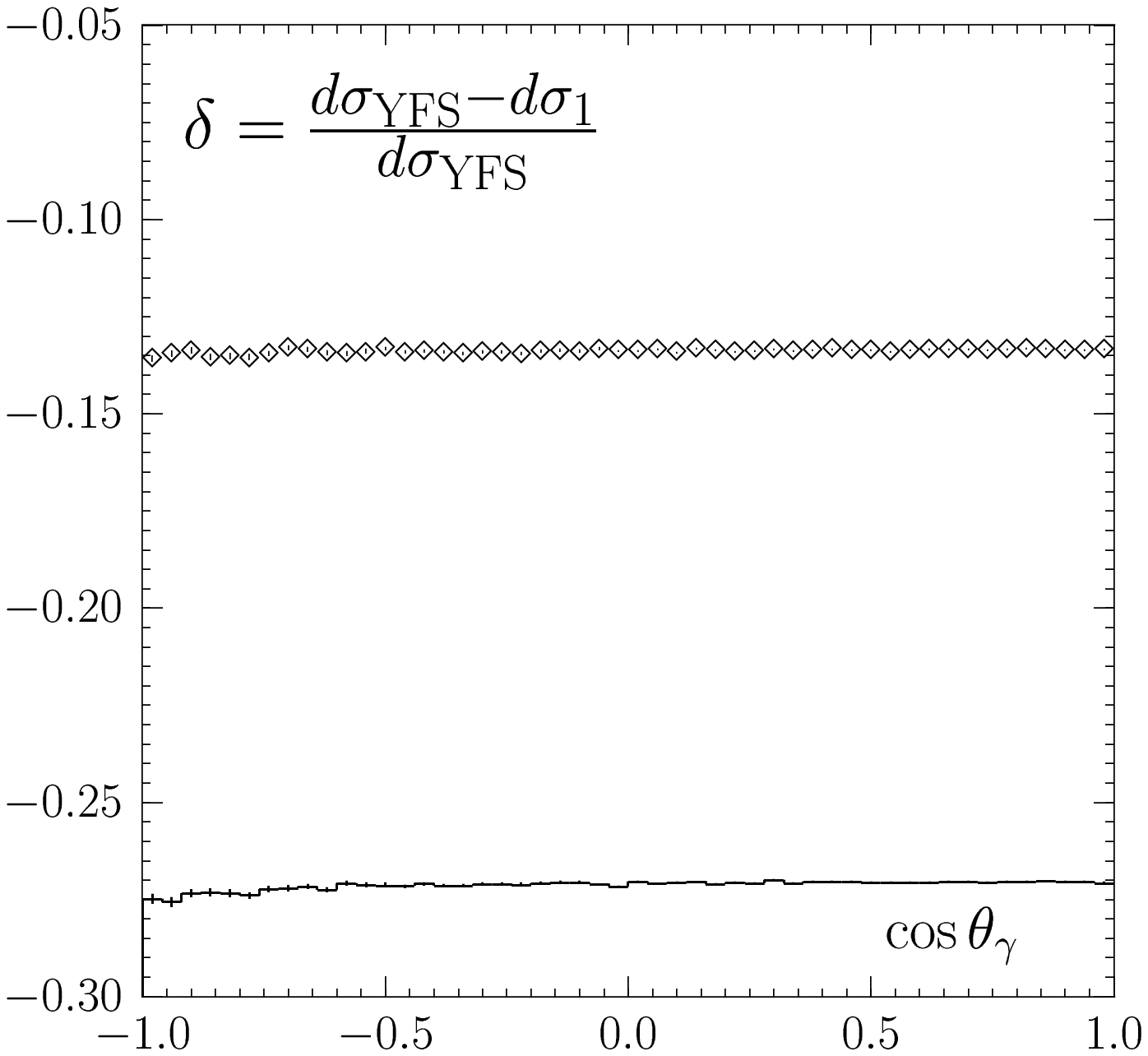            ,width=80mm,height=70mm}
}}
  
\put(800,700){\makebox(0,0)[lb]{
\epsfig{file=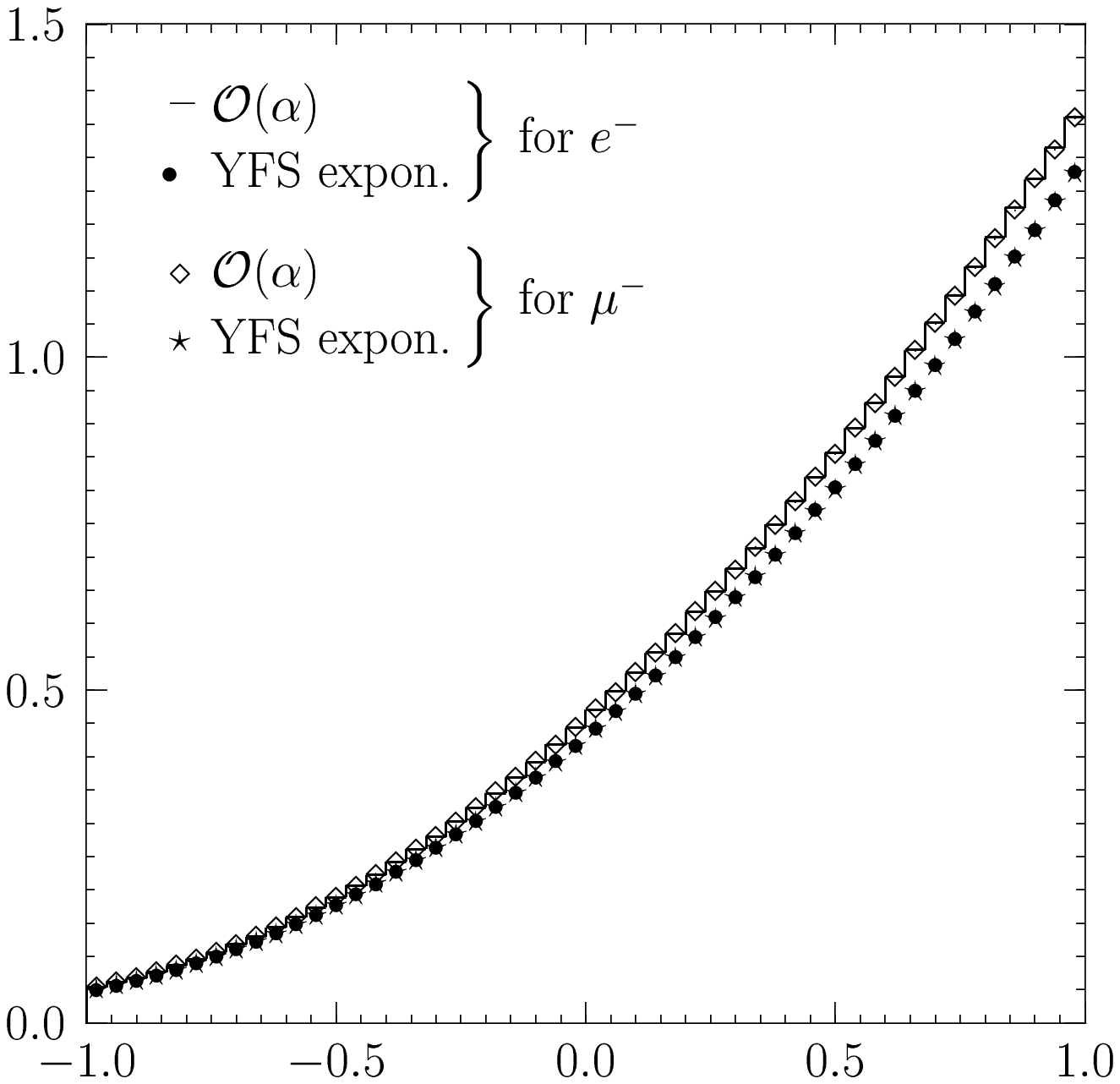                ,width=80mm,height=70mm}
}}
  
\put(800,  0){\makebox(0,0)[lb]{
\epsfig{file=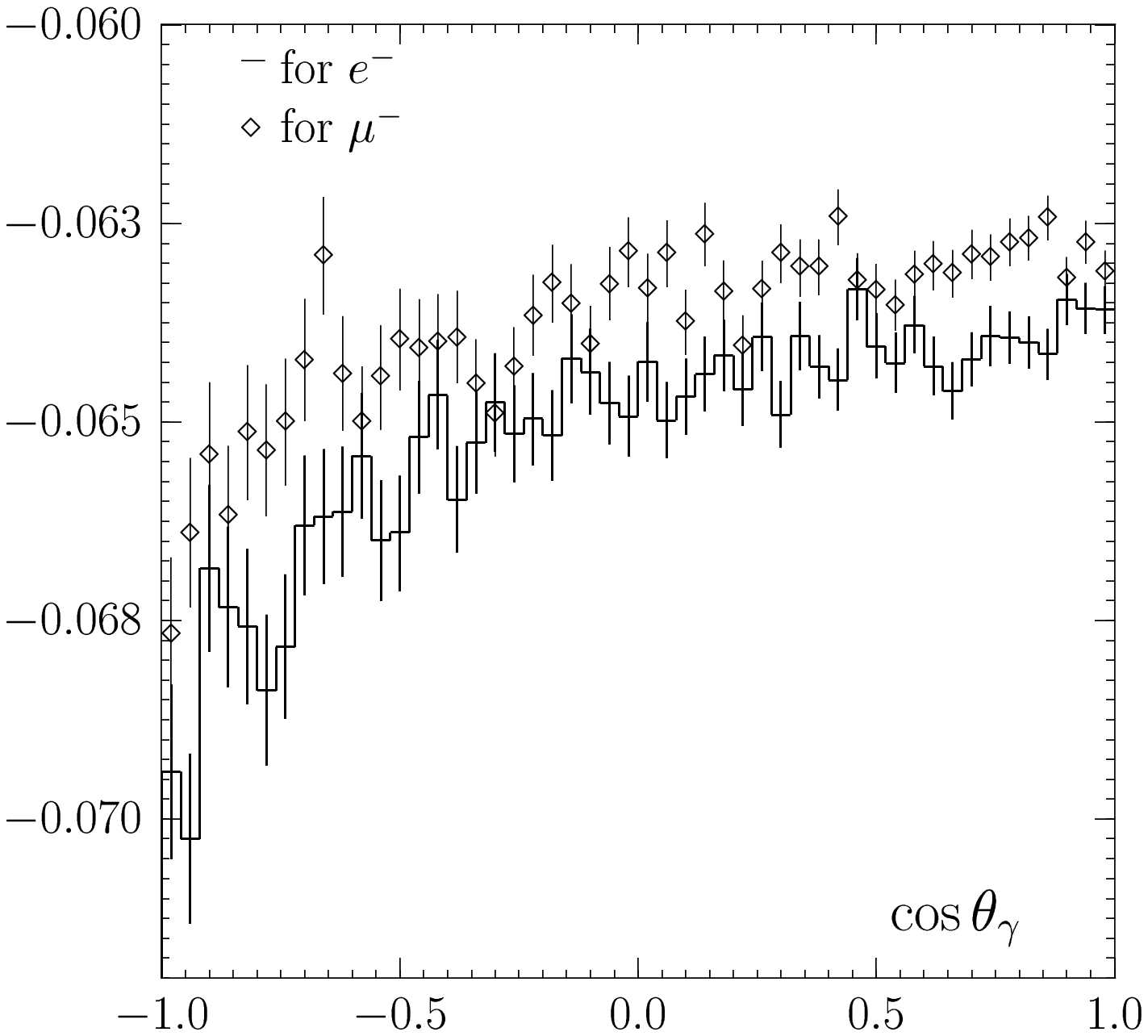            ,width=80mm,height=70mm}
}}
\end{picture}
\caption{\small\sf
 Distributions of the cosine of the hardest-photon polar angle      
 for BARE and CALO acceptances.                                  
}
\label{fig:cg}        
\end{figure}

Here we presents the results from WINHAC
for some distributions at the Born level, with \oal\ QED-like corrections
and including the YFS exponentiation. These results are given for
two kinds of event selection: {\sf BARE} -- where the corresponding observables
are obtained from bare-lepton four-momenta and no cuts are applied,
and {\sf CALO} -- where the photon four-momenta are combined
with the charged-lepton four-momenta if the opening angle between their
directions $\angle(\vec{q}_l,\vec{k})\le 5^{\circ}$; such photons are
discarded. No extra cuts are applied. The {\sf BARE} acceptance is
closer to experimental event selections for muons, while the {\sf CALO}
is closer to the ones for electrons. We, however, use them for both
types of final states.

In Fig.~\ref{fig:El} we present the distributions of the electron and
muon energy for the fixed \oal\ corrections and for the YFS exponentiation.
The upper plots show the absolute distributions for the {\sf BARE} 
and {\sf CALO} acceptances, respectively, while the lower plots show
the relative differences between these two calculations, also 
for the {\sf BARE} and {\sf CALO} acceptances.
At the Born level, the charged-lepton energy is fixed at 
$E_l\approx \frac{1}{2}\sqrt{s}$; therefore, the energy tails for 
$E_l < \frac{1}{2}\sqrt{s}$ in the above plots are the result of the
real-photon radiation in $W$-boson decay. The YFS-exponentiation corrections
beyond the fixed \oal-level are large for the {\sf BARE} acceptance:
up to $\sim 15\%$ for electrons and up to $\sim 8\%$ for muons -- they
differ for these two decay channels. For the {\sf CALO} acceptance
these corrections are smaller, up to $\sim 4\%$, and they are almost
identical for electrons and muons. This is because in this case
the large corrections due to lepton-mass-log terms have been excluded
by the photon--lepton recombination. As can be seen, the largest
(negative) corrections are in the first radiative bin (i.e.\ the second 
highest one in the upper plots) and they change sign for low lepton 
energies.

In Fig.~\ref{fig:Eg} we show the distributions of the hardest photon
energy for the electron and muon $W$-decay channels. The notation
is similar to that in Fig.~\ref{fig:El}. Again, for the {\sf BARE} acceptance
the YFS-exponentiation corrections beyond the fixed \oal-level are
large and different for these two channels: up to $\sim 15\%$ for electrons
and up to $\sim 7\%$ for muons. For {\sf CALO} they are smaller, $\sim 4\%$,
and similar in the two channels. The corrections are largest
for soft photons and decrease with the photon energy.  
  
Finally, in Fig.~\ref{fig:cg} we show the distributions of the {\em cosine} 
of the hardest photon polar angle with respect to the incoming quark 
direction; the notation is as in Fig.~\ref{fig:Eg}. 
For the {\sf BARE} acceptance the YFS-exponentiation corrections beyond the 
fixed \oal-level are large and almost constant: $\sim 28\%$ for electrons
and $\sim 14\%$ for muons. For {\sf CALO} they are smaller, $6$--$7\%$,
and similar in the electron and muon channels. 

As can be seen from Figs.~\ref{fig:El}--\ref{fig:cg}, the YFS exponentiation
affects sizeably radiative events. All the above distributions
have been obtained for the parton-level $W$-boson production
at fixed CMS energy. In the actual proton--(anti)proton collisions
the parton--parton CMS energy can change, which leads to an enhancement
of the FSR corrections, in particular those due to the YFS exponentiation.
This will be investigated elsewhere~\cite{placzek:preparation}.

\section{Summary and outlook}
\label{sec:summary}

In this paper we have presented the calculations of the YFS QED exponentiation
in leptonic $W$-boson decays. We have provided the fully massive spin
amplitudes for the single $W$-boson production and decay, including
the single-real-photon radiation in $W$ decays. We have obtained the
numerically stable representations of the YFS form factor for the 
charged-particle decay. All this has been applied to the process of
Drell--Yan-like $W$-boson production in hadronic collisions 
and implemented, at the parton level, in the Monte Carlo event generator
WINHAC~1.0. For this purpose, an efficient multiphoton MC algorithm
has been developed. The above spin amplitudes have been cross-checked
with the independent analytical representations of the appropriate
matrix elements~\cite{Berends:1985} and they have been found to be in very
good numerical agreement. 
We have also performed several numerical tests of the implemented 
MC algorithm. The results of these tests make us confident in the 
correctness of this MC algorithm. 

Numerically, the YFS-exponentiation corrections beyond the fixed \oal\ 
calculations are at the level of $\sim 10^{-4}$ for the total cross section,
which is the result of the KLN-theorem. However, for some distributions
they can amount to between a few and over $20$ per cent. These corrections
can be significantly reduced when a calorimetric-like recombination
of radiative photons and charged leptons is applied. Such a treatment
is experimentally natural for the electrons in the final state, but less 
obvious for the muons. 

Here we presented the calculations for the QED-like corrections
in the leptonic $W$-boson decays and for the parton-level $W$-production
process only. We are planning to extend this, in the future, to the full 
proton--(anti)proton collisions and to include other \oal\ electroweak
corrections. The next step would be the inclusion of the NLO QCD effects
as well as soft-gluon resummation corrections.
We are also going to perform further tests of the program WINHAC
at the \oal\ and beyond, particularly comparisons with independent
calculations for various observables. Last, but not least,
the full documentation of the MC program WINHAC~\cite{WINHAC:2002} is in 
preparation (to be submitted to {\em Computer Physics Communications}). 
There, the details of the corresponding MC algorithm will be given. 

In this paper, we applied the QED YFS exponentiation in leptonic 
$W$-boson decays to the single-$W$ production process at hadron colliders. 
However, it can also be used to describe the photon radiation in $W$ decays 
in the processes of $W$-pair production at both hadron and electron--positron 
colliders.
In particular, it can be rather easily implemented in our MC event generator 
YFSWW~\cite{yfsww3:2001} for $W^+W^-$ production 
in $e^+e^-$ collisions, which will be necessary for the future linear 
colliders~\cite{TESLA-TDR:2001}.

\vspace{5mm}
\noindent
{\large\bf Acknowledgements}
\vspace{3mm}

\noindent
We would like to thank D.~Bardin and B.F.L.~Ward for useful discussions.
We acknowledge the kind support of the CERN TH and EP Divisions.

\appendix

\section{Spinorial string functions}

Here we provide explicit formulae for the spinorial string functions
introduced in Section~\ref{sec:sa}. The general such function
in the two-component Weyl-spinor basis reads~\cite{hagiwara:1986}
\begin{equation}
S(p_i,a_1,\ldots,a_n,p_j)_{\sigma_i\sigma_j}^{\alpha} =
\chi_{\sigma_i}^{\dagger}(p_i)\left[a_1,\ldots,a_n\right]^{\alpha}
\chi_{\sigma_j}(p_j), 
\label{eq:ssf}
\end{equation}
where
\begin{equation}
\chi_{+}(p) = \frac{1}{\sqrt{2|\vec{p}|(|\vec{p}| + p^3)}}
 \genfrac{\lbrack}{\rbrack}{0pt}{}{|\vec{p}| + p^3}{p^1 + ip^2},
\:\:\:\:\:\:\:
\chi_{-}(p) = \frac{1}{\sqrt{2|\vec{p}|(|\vec{p}| + p^3)}}
 \genfrac{\lbrack}{\rbrack}{0pt}{}{-p^1 + ip^2}{|\vec{p}| + p^3}
\label{eq:chis}
\end{equation}
are the two-component Pauli spinors corresponding to an external
fermion with four-momentum $p=(p^0,\vec{p})=(p^0,p^1,p^2,p^3)$; 
for $p^3=-|\vec{p}|$ we choose
\begin{equation}
\chi_{+}(p) = \genfrac{\lbrack}{\rbrack}{0pt}{}{0}{1},
\:\:\:\:\:\:\:
\chi_{-}(p) = \genfrac{\lbrack}{\rbrack}{0pt}{}{-1}{\:\:\:0}.
\label{eq:chiz}
\end{equation}
The internal part of the above string function 
\begin{equation}
\left[a_1,\ldots,a_n\right]^{\alpha} = 
(\rlap{/}a_1)_{\alpha} (\rlap{/}a_2)_{-\alpha}\ldots 
(\rlap{/}a_n)_{(-1)^{n+1}\alpha}
\label{eq:amat}
\end{equation}
is the product of 2$\times$2 $c$-number matrices,
where 
\begin{equation}
(\rlap{/}a)_{\pm} = 
  \left[  
    \begin{array}{cc}
      a^0 \mp a^3 & \mp (a^1 - ia^2)\\
     \mp (a^1 - ia^2) & a^0 \pm a^3 \\
    \end{array}
   \right]
\label{eq:aslash}
\end{equation}
with $a=(a^0,a^1,a^2,a^3)$ the four-vector in the Minkowski space.

As can be seen, the spinorial function $S$ can be easily evaluated 
numerically for arbitrary $n$. One can just compute a product of internal
2$\times$2 matrices $(\rlap{/}a_i)_{\alpha}$, 
and then multiply the resulting matrix 
by the external 2-dimensional $c$-number vectors $\chi$. 
However, the numerical evaluation
is more efficient if, instead of matrix-by-matrix multiplication, one
performs matrix-by-vector multiplication. In our computation of the
function $S$, we start from multiplying the left-hand-side vector
$\chi^{\dagger}$ by the matrix $(\rlap{/}a_1)_{\alpha}$, and continue
by multiplying the resulting vectors by the consecutive matrices
$(\rlap{/}a_i)_{\alpha}$ until we reach the last matrix, 
$(\rlap{/}a_n)_{\alpha}$. The computation is completed by performing the scalar
product of the final vector of the above multiplication with 
the right-hand-side vector $\chi$.

Three polarization vectors of a massive vector-boson with four-momentum
$k=(k^0,\vec{k})=(k^0,k^1,k^2,k^3)$ and the mass $m$ are, 
in the Cartesian basis,  given by
\begin{equation}
\begin{aligned}
\epsilon^{\mu}(k,\lambda=1) &= \frac{1}{|\vec{k}|k_T}\,
                                \left(0,k^1k^3,k^2k^3,-k_T^2\right),\\
\epsilon^{\mu}(k,\lambda=2) &= \frac{1}{k_T}\, \left(0,-k^2,k^1,0\right),\\
\epsilon^{\mu}(k,\lambda=3) &= \frac{k^0}{m|\vec{k}|}\,
                              \left(\frac{|\vec{k}|^2}{k^0},k^1,k^2,k^3\right),
\end{aligned}
\label{eq:polvec}
\end{equation}
where $k_T=\sqrt{(k^1)^2 + (k^2)^2}$ is the transverse momentum.
For massless vector bosons, such as photons, $\epsilon^{\mu}(\lambda=3)=0$,
i.e. there are only two non-zero polarizations 
$\epsilon^{\mu}(\lambda=1)$ and $\epsilon^{\mu}(\lambda=2)$. 
Helicity eigenstates can be obtained from the above polarization vectors
through
\begin{equation}
\begin{aligned}
  \epsilon_{hel}(k,\lambda=\pm) &= \frac{1}{\sqrt{2}}\,
    \left[\,\mp\epsilon(k,\lambda=1)
    - i\epsilon(k,\lambda=2)\,\right]\,, \\
  \epsilon_{hel}(k,\lambda=0) &= \epsilon(k,\lambda=3).
\end{aligned}
\label{eq:epshel}
\end{equation}

\section{The YFS IR functions} 

In Ref.~\cite{yfsww2:1996} we provided the general formulae for
the YFS IR functions $\Re B$ and $\tilde{B}$ for a pair of 
charged particles of arbitrary masses and four-momenta. This representation
works very well for particle production or scattering processes;
however, it becomes numerically unstable for charged-particle decays. 
Therefore, for the process  
\begin{equation}
W^{\pm}(Q) \longrightarrow l^\pm(q) + \overset{(-)}{\nu_l}(q'),
\label{ap:wdec}
\end{equation}
we have to obtain different representations for these functions. 
A specific feature of the above process is that 
a QED-radiation dipole is stretched between a decaying particle 
($W$) and a decay product ($l$). For such a dipole, a four-momentum transfer
between its constituents is non-negative:
\begin{equation}
t = (Q-q)^2 \geq 0,
\label{eq:tran}
\end{equation}
in contrast to scattering processes. In such a case, the corresponding
IR integrals have to be calculated in a slightly different way than
was done in Ref.~\cite{yfsww2:1996} for $t<0$. 
Special care is also needed for the limiting case $t = 0$, 
which occurs for the two-body 
$W$-boson leptonic decay when the neutrino mass is neglected.
For the sake of numerical stability, it has to be treated separately.

The YFS virtual- and real-photon IR functions for a pair of charged particles 
with the four-momenta $(Q,\,q)$ are defined as follows~\cite{yfs:1961}
\begin{align}
  B(Q,q;m_{\gamma}) & = \frac{i}{8\pi^3} 
   \int \frac{d^4k}{k^2 - m_{\gamma}^2 + i\varepsilon}
     \left(\frac{2q - k}{k^2 - 2kq + i\varepsilon} -
           \frac{2Q - k}{k^2 - 2kQ + i\varepsilon}\right)^2,
\label{eq:IRfunv}\\
 \tilde{B}(Q,q;m_{\gamma},k_s) & = -\frac{1}{8\pi^2}
   \int_{k^0< k_s}\frac{d^3k}{k^0}\left(\frac{q}{kq} - \frac{Q}{kQ}\right)^2, 
\label{eq:IRfunr} 
\end{align} 
where $m_{\gamma}$ is a dummy photon mass used 
to regularize the IR-divergent integrals ($m_{\gamma}\ll k_s$), while
$k_s$ is the soft-photon cut-off, up to which the integration over
the real-photon four-momenta is carried over analytically ($k_s\ll Q^0$). 
The explicit analytical formulae for these functions are presented below.

\subsection{The virtual-photon IR function}  

The virtual-photon IR function reads as follows:
\begin{description}
\item[I.] \underline{$t=(Q-q)^2 > 0$:}
\begin{equation}
2\alpha\Re B(Q,q;m_{\gamma}) = \frac{\alpha}{\pi}\left\{ 
  \left[\,\nu A(Q,q) - 1\, \right]\,\ln\frac{m_{\gamma}^2}{Mm} +
        \frac{1}{2} A_1(Q,q) - \nu A_3(Q,q)\right\},
\label{eq:ReBtpos}
\end{equation}
with
\begin{align}
A(Q,q) & = \frac{1}{\lambda}\ln\frac{\lambda+\nu}{Mm}, 
\label{eq:Afun} \\
A_1(Q,q) & = \frac{M^2-m^2}{t}\ln\frac{M}{m} - \frac{2\lambda^2}{t}\,A(Q,q) -2,
\label{eq:A1fun} \\
A_3(Q,q) & = A(Q,q)\,\ln\frac{2\lambda}{Mm} + \frac{1}{\lambda}\bigg[
             \frac{1}{4}\left(\ln\frac{\lambda+\nu}{M^2} + 
                     2\ln\frac{\lambda - \nu + M^2}{t}\right)
              \ln\frac{\lambda+\nu}{M^2} \notag\\
         & + \frac{1}{4}\left(\ln\frac{\lambda+\nu}{m^2} -
                     2\ln\frac{\lambda + \nu - m^2}{m^2}\right)  
              \ln\frac{\lambda+\nu}{m^2} 
        \notag\\
        &  + \frac{1}{2}\ln\eta\ln(1+\eta) - \frac{1}{2}\ln\zeta\ln(1+\zeta)
         + \Re\mathrm{Li}_2(-\eta) - \Re\mathrm{Li}_2(-\zeta)\bigg],
\label{eq:A3fun}
\end{align} 
where
\begin{equation}
\begin{aligned}
\,& \nu = Qq, \:\:\:\: \lambda=\sqrt{(\nu-Mm)(\nu+Mm)}, \:\:\:\:
  Q^2 = M^2,\:\: q^2 = m^2,\:\: M>m\,, \\
\,& t = M^2 + m^2 - 2\nu, \:\:\:\:\:\:\:\: 
    Mm\leq \nu < \frac{1}{2}\left(M^2 + m^2\right)\,, \\
\,& \eta = \frac{m^2t}{2\lambda(2\lambda+\nu-m^2)}, \:\:\:\:\:\:\:\:
  \zeta = \frac{\lambda+\nu}{m^2}\,\eta\,,
\label{eq:ReBnot}
\end{aligned}
\end{equation}
and
\begin{equation}
\mathrm{Li}_2(y) = -\int_0^y\frac{dx}{x} \ln(1-x),\:\:\:\:\:\:\:\:
                    |\arg(1-y)|< \pi,
\label{eq:dilog}
\end{equation}
is the Spence dilogarithm function.

\item[II.] \underline{$t=(Q-q)^2 = 0$:}
\begin{equation}
2\alpha\Re B(Q,q;m_{\gamma}) = \frac{\alpha}{\pi}\left\{
  \left(\frac{M^2+m^2}{M^2-m^2}\ln\frac{M}{m} - 1 \right)
   \left(\ln\frac{m_{\gamma}^2}{Mm}+ \frac{1}{2} \right)\right\}\,.
\label{eq:ReBtzer}
\end{equation}
In the limit $m\ll M$ we get
\begin{equation}
2\alpha\Re B(Q,q;m_{\gamma})\underset{m\ll M}{=} \frac{\alpha}{\pi}
  \left\{2 \left( \ln\frac{M}{m} - 1 \right)\ln\frac{m_{\gamma}}{M}
   + \ln^2\frac{M}{m} -  \frac{1}{2} \ln\frac{M}{m} - \frac{1}{2} \right\}\,.
\label{eq:ReBtzsm}
\end{equation}

\end{description}

\subsection{The real-photon IR function}  

For the real-photon IR function we obtain
\begin{description}
\item[I.] \underline{$t=(Q-q)^2 > 0$:}
\begin{equation}
\begin{aligned}
\tilde{B}(Q,q;m_{\gamma},k_s) = \frac{\alpha}{\pi}\,\bigg\{&  
  \left[\,\nu A(Q,q) - 1\, \right]\,\ln\frac{4k_s^2}{m_{\gamma}^2} -
        \frac{M^2}{2} A_4(Q,Q) - \frac{m^2}{2} A_4(q,q) \\
  & - \nu A_4(Q,q)\bigg\},
\end{aligned}
\label{eq:Bttpos}
\end{equation}
with 
\begin{align}
A_4(p,p) & = \frac{1}{p^2\beta}\ln\frac{1-\beta}{1+\beta},\:\:\:\:\:\:\:\: 
             \beta =\frac{|\vec{p}|}{p^0},
 \label{eq:A4pp} \\
A_4(Q,q) & = \frac{1}{\kappa}\left\{ \ln\left|\frac{V^2}{t}\right|
              \sum_{i=0}^{1} (-1)^{n+1}
     \left[\,X(z_i;y_1,y_4,y_2,y_3) + R(z_i)\,\right]
    \right\}\,,
\label{eq:A4fun}
\end{align} 
where
\begin{equation}
\begin{aligned}
\,& R(z)  = Y_{14}(z) + Y_{21}(z) + Y_{32}(z) - Y_{34}(z) \\
\,&\hspace{10mm} + \frac{1}{2}X(z;y_1,y_2,y_3,y_4)X(z;y_2,y_3,y_1,y_4)\,,\\
\,& Y_{ij}(z)  = 2Z_{ij}(z) + 
           \frac{1}{2}\ln^2\left|\frac{z-y_i}{z-y_j}\right|\,,\\
\,& Z_{ij}(z) = \Re \mathrm{Li}_2\left(\frac{y_j-y_i}{z-y_i}\right)\,,\\
\,& X(z;a,b,c,d) =\ln\left|\frac{(z-a)(z-b)}{(z-c)(z-d)}\right|\,,
\label{eq:RXYZ}
\end{aligned}
\end{equation}
and
\begin{equation}
\begin{aligned}
\,& z_0 = \frac{|\vec{q}|}{T}\,, \hspace{49mm} 
  z_1 = \frac{|\vec{Q}|}{T} - 1\,; \\
\,& y_1 = -\frac{1}{2T}\left[\,T + \Omega - 
         \frac{\omega\delta + \kappa}{t}\,V\,\right], \hspace{5mm} 
  y_2 = y_1 - \frac{\kappa V}{tT}\,, \\
\,& y_3 = -\frac{1}{2T}\left[\,T - \Omega +
         \frac{\omega\delta + \kappa}{V}\,\right], \hspace{9mm}
  y_4 = y_3 + \frac{\kappa}{TV}\,; \\
\,& \kappa =\sqrt{(\omega^2 - t)(\delta^2 - t)}\,,
  \hspace{5mm} \delta = M - m, \:\: \omega = M + m\,, \\
\,& T = \sqrt{\Delta^2 - t}\,, \:\: V = \Delta + T, \hspace{5mm}
  \Delta = Q^0 - q^0, \:\: \Omega = Q^0 + q^0\,,
\label{eq:Btnot}
\end{aligned}
\end{equation}
while $\nu$ and $A(Q,q)$ are as given in the previous subsection.
We have checked that this analytical representation is numerically stable 
for $t \gtrsim 10^{-10}\,\mathrm{GeV}^2$, when computed in any Lorentz
frame, which is neither $W$ nor $l$ rest frame. In these frames we need
an explicit analytical formula for $A_4(p,p)$ in the limit $\beta\rightarrow 
0$. 
It reads
\begin{equation}
A_4(p,p) \underset{p=(m,\vec{0})}{=} -\frac{2}{m^2}\,.
\label{eq:A4pprf}
\end{equation}
In the $W$ rest frame, i.e. for $Q=(M,\vec{0})$,  the function
$A_4(Q,q)$ can be simplified to get
\begin{equation}
\begin{aligned}
A_4(Q,q)  
           {=}
          \frac{1}{2M\bar{q}}\bigg[ &
          \ln\frac{M-E+\bar{q}}{M-E-\bar{q}}\ln\frac{E+\bar{q}}{M}
         - 2\ln\frac{2\bar{q}(M-E+\bar{q})}{Mm}\ln\frac{E+\bar{q}}{M} \\
       + & 2\Re\mathrm{Li}_2\left(\frac{E-\bar{q}}{M}\right)  
         - 2\Re\mathrm{Li}_2\left(\frac{E+\bar{q}}{M}\right) \\
       + & \Re\mathrm{Li}_2\left(\frac{M-E-\bar{q}}{-2\bar{q}}\right)  
         - \Re\mathrm{Li}_2\left(\frac{M-E+\bar{q}}{2\bar{q}}\right) \\ 
       + & \Re\mathrm{Li}_2\left(\frac{M(E+\bar{q})-m^2}{2M\bar{q}}\right)  
         - \Re\mathrm{Li}_2\left(\frac{M(E-\bar{q})-m^2}{-2M\bar{q}}\right)
        \,\bigg]\,,  
\end{aligned}
\label{eq:A4Qqrf}
\end{equation} 
where $E=q^0,\,\bar{q}=|\vec{q}|$.

\item[I.] \underline{$t=(Q-q)^2 = 0$:}

For $t=0$ the functions $A(Q,q), A_4(Q,Q), A_4(q,q)$ can remain the same
as for $t>0$, but we need a new, numerically stable, representation for
the function $A_4(Q,q)$. It can be cast in the form
\begin{equation}
A_4(Q,q) = \frac{1}{\mu^2}\left[\,\ln\frac{2\Delta^2}{\mu^2}
                                  \ln\left|\frac{\xi_2\xi_3}{\xi_1}\right|
             + U(z_1) - U(z_0)\right]\,,
\label{ea:A4tz}
\end{equation}  
where
\begin{equation}
\begin{aligned}
\,& U(z) = \frac{1}{2}\ln^2\left|\frac{(z - y_1)(z-y_2)}{z - y_3}\right|
       + \ln|z-y_1| \ln\frac{|z-y_1|}{(z-y_2)^2} \\
\,&\hspace{10mm} + 2\Re \mathrm{Li}_2\left(\frac{y_2-y_1}{z-y_1}\right)
               + 2\Re \mathrm{Li}_2\left(\frac{y_3-y_2}{z-y_2}\right)\,;\\
\,& \xi_i = \frac{z_0 - y_i}{z_1 - y_i}\,, \hspace{5mm}
  z_0 = \frac{|\vec{q}|}{\Delta}\,,\:\:\:\: 
  z_1 = \frac{|\vec{Q}|}{\Delta} - 1\,;\\
\,& y_1 = \frac{q^0}{\Delta}\,,\:\:\:\: y_2 = y_1 - \frac{\mu^2}{2\Delta^2}\,,
  \:\:\:\: y_3 = - y_1 + \frac{2m^2}{\mu^2}\,; \\
\,& \Delta = Q^0 - q^0\,, \:\:\:\: \mu^2 = M^2 - m^2\,.
\label{eq:A4tznot}
\end{aligned}
\end{equation}
In the $W$ rest frame we get 
\begin{equation}
A_4(Q,q) \underset{Q=(M,\vec{0})}{=} -\frac{2}{M^2-m^2}
          \left[\,\ln^2\frac{M}{m} 
        + \mathrm{Li}_2\left(\frac{M^2-m^2}{M^2}\right) \,\right]\,.
\label{eq:A4Qqtzrf}
\end{equation}
Then, in the small-lepton-mass limit, $m\ll M$, we obtain a simple expression
for the function $\tilde{B}$ in the $W$ rest frame:
\begin{equation}
2\alpha\tilde{B}(Q,q;m_{\gamma},k_s) \underset{m\ll M}{=}
                  \frac{\alpha}{\pi}\left\{2\left(\ln\frac{M}{m}-1\right)
                       \ln\frac{2k_s}{m_{\gamma}} -\ln^2\frac{M}{m}
                      +\ln\frac{M}{m} + 1 -\frac{\pi^2}{6}\right\}.
\label{eq:Btrfsma}
\end{equation}
After combining this with the virtual-photon function 
of Eq.~(\ref{eq:ReBtzsm}),
we obtain a simple expression for the YFS form factor in the $W$ rest frame: 
\begin{equation}
Y(Q,q;k_s) \underset{m\ll M}{=}
        \frac{\alpha}{\pi}\left\{2\left(\ln\frac{M}{m}-1\right)
         \ln\frac{2k_s}{M} + \frac{1}{2}\ln\frac{M}{m} 
         - \frac{1}{2} -\frac{\pi^2}{6}\right\}.
\label{eq:Yrfsma}
\end{equation}
As can be seen see explicitly, 
it is free of the IR singularity as well as of the Sudakov double-logarithms.
\end{description} 


\end{document}